\documentclass[12pt, a4paper]{article}
\usepackage{float}
\usepackage{graphicx}
\usepackage{amstext}
\usepackage[fleqn]{amsmath}
\usepackage{amssymb}	% Extra maths symbols
\usepackage{url}
\usepackage{tabularx}
\usepackage{mathtools}
\usepackage{mathrsfs}
\usepackage{color}
\usepackage{graphics}
\usepackage{multirow}
\usepackage{soul}
\usepackage{booktabs} 
\usepackage{colortbl} 
\usepackage{xfrac}
\usepackage{xspace}
\usepackage{verbatim}
\usepackage{longtable}
\usepackage{lineno}
\usepackage{rotating}
\usepackage{lastpage} % Used to determine the number of pages in the document (for "Page X of Total")
\usepackage[english]{babel} % English language hyphenation
\usepackage{microtype} % Better typography
\usepackage[svgnames]{xcolor}
\usepackage[small, labelfont=bf, up]{caption} % Custom captions under/above tables and figures
\usepackage{enumitem} % Required for customising lists
\setlist{noitemsep} % Remove spacing between bullet/numbered list elements
\usepackage{sectsty} % Enables custom section titles
\allsectionsfont{\usefont{OT1}{phv}{b}{n}} % Change the font of all section commands (Helvetica)
\usepackage[raggedright]{titlesec}
\usepackage{multicol} 
\usepackage[superscript,biblabel]{cite}
\usepackage{hyperref}
\hypersetup{colorlinks=true,linkcolor=Blue,citecolor=Blue,filecolor=Blue,urlcolor=Blue}

\usepackage{geometry} % Required for adjusting page dimensions

\usepackage{times}
\usepackage[T1]{fontenc} % Output font encoding for international characters
\usepackage[utf8]{inputenc} % Required for inputting international characters

\usepackage{fancyhdr} % Needed to define custom headers/footers
\fancypagestyle{firstpage}{ % Page style for the first page with the title
	\fancyhf{}
}

\newcommand{\authorstyle}[1]{{\normalsize\usefont{OT1}{phv}{b}{n}\color{Black}#1}} % Authors style (Helvetica)

\newcommand{\institution}[1]{{\scriptsize\usefont{OT1}{phv}{m}{n}\color{Black}#1}} % Institutions style (Helvetica)

\usepackage{titling} % Allows custom title configuration

\newcommand{\HorRule}{\color{DarkBlue}\rule{\linewidth}{1pt}} % Defines the gold horizontal rule around the title

\pretitle{
	\vspace{-20pt} % Move the entire title section up
	\noindent\HorRule\vspace{5pt} % Horizontal rule before the title
	\fontsize{22}{26}\usefont{OT1}{phv}{b}{n}\selectfont % Helvetica
	\color{DarkBlue} % Text colour for the title and author(s)
}

\posttitle{\par\vskip 5pt} % Whitespace under the title

\preauthor{\vspace{0pt}} % Anything that will appear before \author is printed

\postauthor{ % Anything that will appear after \author is printed
	\vspace{5pt} % Space before the rule
	\par\noindent\HorRule % Horizontal rule after the title
	\vspace{0pt} % Space after the title section
}

\newcommand{\msun}{\mbox{M$_\odot$}}

\newcommand{\pc}{\mbox{${\rm pc}$}}

\newcommand{\kms}{\mbox{${\rm km}~{\rm s}^{-1}$}}

\newcommand{\be}{\begin{equation}}
\newcommand{\ee}{\end{equation}}
\newcommand*\Bell{\ensuremath{\boldsymbol\ell}}
\newcommand{\ppv}{\mbox{$p$-$p$-$v$}}
\newcommand{\farc}{\overset{\prime\prime}{.}}
\newcommand{\fdg}{\overset{^\circ}{.}}

\title{Ubiquitous velocity fluctuations throughout \\
the molecular interstellar medium}
\author{\noindent
    \authorstyle{
        Jonathan~D.~Henshaw\textsuperscript{1*},
        J.~M.~Diederik Kruijssen\textsuperscript{2},
        Steven~N.~Longmore\textsuperscript{3}, \\
        Manuel~Riener\textsuperscript{1},
        Adam~K.~Leroy\textsuperscript{4},
        Erik~Rosolowsky\textsuperscript{5},
        Adam~Ginsburg\textsuperscript{6}, \\
        Cara~Battersby\textsuperscript{7}, 
        M\'elanie Chevance\textsuperscript{2}, 
        Sharon~E.~Meidt\textsuperscript{8}, 
        Simon~C.~O.~Glover\textsuperscript{9}, \\
        Annie~Hughes\textsuperscript{10,11},
        Jouni~Kainulainen\textsuperscript{12}, 
        Ralf~S.~Klessen\textsuperscript{9}, 
        Eva Schinnerer\textsuperscript{1}, \\
        Andreas Schruba\textsuperscript{13},
        Henrik~Beuther\textsuperscript{1}, 
        Frank~Bigiel\textsuperscript{14}, 
        Guillermo~A.~Blanc\textsuperscript{15,16}, \\
        Eric~Emsellem\textsuperscript{17,18}, 
        Thomas Henning\textsuperscript{1}, 
        Cynthia~N.~Herrera\textsuperscript{19}, 
        Eric~W.~Koch\textsuperscript{5}, \\
        J\'er\^ome~Pety\textsuperscript{19,20},
        Sarah~E.~Ragan\textsuperscript{21}, 
        and Jiayi~Sun\textsuperscript{4}}
    \newline\newline
    \textsuperscript{1}\institution{Max Planck Institut f\"{u}r Astronomie, K\"{o}nigstuhl 17, D-69117 Heidelberg, Germany}\\
    \textsuperscript{2}\institution{Astronomisches Rechen-Institut, Zentrum f{\"u}r Astronomie der Universit{\"a}t Heidelberg, M\"{o}nchhofstr. 12-14, 69120 Heidelberg, Germany}\\
    \textsuperscript{3}\institution{Astrophysics Research Institute, Liverpool John Moores University, IC2, Liverpool Science Park, 146 Brownlow Hill, Liverpool L3 5RF, United Kingdom}\\
    \textsuperscript{4}\institution{Department of Astronomy, The Ohio State University, 4055 McPherson Laboratory, 140 West 18th Avenue, Columbus, OH 43210, USA}\\
    \textsuperscript{5}\institution{Department of Physics, University of Alberta, Edmonton, AB, Canada}\\
    \textsuperscript{6}\institution{Department of Astronomy, University of Florida, Bryant Space Science Center, Gainesville FL 32611, USA}\\
    \textsuperscript{7}\institution{Department of Physics, 196 Auditorium Road, Unit 3046, University of Connecticut, Storrs, CT 06269-3046, USA}\\
    \textsuperscript{8}\institution{Sterrenkundig Observatorium, Universiteit Gent, Krijgslaan 281 S9, B-9000 Gent, Belgium}\\
    \textsuperscript{9}\institution{Institut f{\"u}r Theoretische Astrophysik, Zentrum f{\"u}r Astronomie der Universit{\"a}t Heidelberg, Albert-Ueberle-Str. 2, 69120 Heidelberg, Germany}\\
    \textsuperscript{10}\institution{Universit\'{e} de Toulouse, UPS-OMP, IRAP, F-31028 Toulouse cedex 4, France}\\
    \textsuperscript{11}\institution{CNRS, IRAP, 9 av. du Colonel Roche, BP 44346, F-31028 Toulouse cedex 4, France}\\
    \textsuperscript{12}\institution{Chalmers University of Technology, Department of Space, Earth and Environment, SE-412 93 Gothenburg, Sweden}\\
    \textsuperscript{13}\institution{Max Planck Institut f\"{u}r Extraterrestrische Physik, Giessenbachstra{\ss}e 1, D--85748, Garching bei M\"unchen, Germany}\\
    \textsuperscript{14}\institution{Argelander-Institut f\"{u}r Astronomie, Universit\"{a}t Bonn, Auf dem H\"{u}gel 71, 53121 Bonn, Germany}\\
    \textsuperscript{15}\institution{Observatories of the Carnegie Institution for Science, 813 Santa Barbara Street, Pasadena, CA 91101, USA}\\
    \textsuperscript{16}\institution{Departamento de Astronomía, Universidad de Chile, Casilla 36-D, Santiago, Chile}\\
    \textsuperscript{17}\institution{European Southern Observatory, Karl-Schwarzschild Stra{\ss}e 2, D--85748 Garching bei M\"{u}nchen, Germany}\\
    \textsuperscript{18}\institution{Univ Lyon, Univ Lyon1, ENS de Lyon, CNRS, Centre de Recherche Astrophysique de Lyon UMR5574, F-69230 Saint-Genis-Laval France}\\
    \textsuperscript{19}\institution{IRAM, 300 rue de la Piscine, 38406 Saint-Martin-d'H\'eres, France}\\
    \textsuperscript{20}\institution{LERMA, Observatoire de Paris, PSL Research University, CNRS, Sorbonne Universites, Univ. Paris 06, 75005 Paris, France}\\
    \textsuperscript{21}\institution{School of Physics \& Astronomy, Cardiff University, Queen's Building, The Parade, Cardiff, CF24 3AA, UK}
}

\date{}
\begin{document}
%%TC:ignore
\maketitle
%%TC:endignore

\newpage 
\vspace{-20pt}

%%%%%%%%%%%%%%%%%%%%%%%%%%%%%%%%%%%%%%%%%%%%%%%%%%%%%%%%%%%%%%%%%%%%%%%%%%%%%%%%%%%%%%%%%%%%%%%%
%%%%%%%%%%%%%%%%%%%%%%%%%%%%%%%%%%          Main Text          %%%%%%%%%%%%%%%%%%%%%%%%%%%%%%%%%
%%%%%%%%%%%%%%%%%%%%%%%%%%%%%%%%%%%%%%%%%%%%%%%%%%%%%%%%%%%%%%%%%%%%%%%%%%%%%%%%%%%%%%%%%%%%%%%%

\textbf{The density structure of the interstellar medium (ISM) determines where stars form and release energy, momentum, and heavy elements, driving galaxy evolution\cite{mckee07,zinnecker07,naab17,kruijssen19}. Density variations are seeded and amplified by gas motion, but the exact nature of this motion is unknown across spatial scale and galactic environment\cite{dobbs14}. Although dense star-forming gas likely emerges from a combination of instabilities\cite{elmegreen83,kim03}, convergent flows\cite{vazquez-semadeni19}, and turbulence\cite{padoan14}, establishing the precise origin is challenging because it requires quantifying gas motion over many orders of magnitude in spatial scale. Here we measure\cite{henshaw16,henshaw19,riener19} the motion of molecular gas in the Milky Way and in nearby galaxy NGC\,4321, assembling observations that span an unprecedented spatial dynamic range ($10^{-1}{-}10^3$~pc). We detect ubiquitous velocity fluctuations across all spatial scales and galactic environments. Statistical analysis of these fluctuations indicates how star-forming gas is assembled. We discover oscillatory gas flows with wavelengths ranging from $0.3{-}400$~pc. These flows are coupled to regularly-spaced density enhancements that likely form via gravitational instabilities\cite{henshaw16c, elmegreen18}. We also identify stochastic and scale-free velocity and density fluctuations, consistent with the structure generated in turbulent flows\cite{padoan14}. Our results demonstrate that ISM structure cannot be considered in isolation. Instead, its formation and evolution is controlled by nested, interdependent flows of matter covering many orders of magnitude in spatial scale.}

%%TC:ignore

We use observations that trace molecular gas in a variety of galactic environments and span a wide range of spatial scales. We measure the position-position-velocity ($p$-$p$-$v$) structure of the molecular ISM from 0.1\,pc scales, relevant for individual star-forming cores, up to the scales of individual giant molecular clouds (GMCs), now accessible in nearby galaxies using facilities such as the Atacama Large Millimeter/submillimeter Array (ALMA). On large (from 100\,pc to $>$1000\,pc) scales, we analyse observations of nearby galaxy NGC\,4321 from the Physics at High Angular resolution in Nearby GalaxieS (PHANGS-ALMA) survey. At intermediate (from 1\,pc to 100\,pc) scales, we target both the Galactic disc and the Central Molecular Zone (CMZ, i.e.\ the central $500$\,pc) of the Milky Way with data from the Galactic Ring Survey\cite{jackson06} and the Mopra CMZ survey\cite{jones12}, respectively. On small scales (from 0.1\,pc to around 10\,pc) we include observations of two dense molecular clouds, G035.39--00.33\cite{henshaw14} in the Galactic disc and G0.253+0.016\cite{henshaw19} in the CMZ.  

We extract the kinematics of the gas using spectral decomposition, modelling each spectrum as a set of individual Gaussian emission features\cite{henshaw16,henshaw19,riener19}. Spectral decomposition is advantageous because it yields a description of all prominent emission features observed in spectroscopic data. We visualise the results using the peak intensities and velocity centroids of the modelled emission features (\ref{Figure:wiggles}). This method facilitates the detection of small fluctuations in velocity, which can often be hidden by techniques that either average in the spectral domain or integrate over one of the two spatial directions. We discover striking similarity in the $p$-$p$-$v$ structure of the molecular ISM in all of our selected environments, despite our observations probing vastly different spatial scales. The $p$-$p$-$v$ volumes presented in \ref{Figure:wiggles} (see also the Supplementary Videos) reveal the complex multi-scale dynamical structure of the molecular ISM. The `wiggles' evident in all data sets represent localised gas flows superposed on larger scale ordered motion. 

Measuring the dynamical coupling between ISM density enhancements and their local environment is key to understanding the formation of hierarchically-structured star-forming gas\cite{mckee07}. The structure of dense gas that is weakly dynamically coupled to the local environment should show little or no correlation with observed gas flows. In contrast, density structures produced by on-going convergent flows should be closely coupled to the gas flows\cite{klessen10, vazquez-semadeni19}. In the specific case of instability-driven structure formation, preferred characteristic scales may be present in both density\cite{elmegreen83,tafalla15} and velocity. Between the scales of energy injection and dissipation, turbulence, by comparison, is characterised by scale-free fluctuations in both density and velocity\cite{padoan14}. 

We select a sub-region from each data set presented in \ref{Figure:wiggles} to represent the hierarchy of the ISM. We first select part of the southernmost dominant spiral arm in NGC\,4321 and, in the CMZ, a portion of the gas orbiting the centre of the Galaxy at a galactocentric radius of about 100\,pc. We then select two GMCs, one in the Galactic disc and another in the CMZ. Finally, we select an individual filament which is embedded within our selected GMC in the Galactic disc. Maps of all the regions can be found in \ref{fig:edfmaps}.

We use structure functions to measure the coupling between the observed gas flows and the physical structure of the gas in each sub-region (see Methods and Supplementary Information). Normalised noise-corrected structure functions of velocity and gas density are presented in \ref{Figure:structfunc}. The velocity structure functions measured in 1-D along the crests of the spiral arm of NGC\,4321 (\ref{Figure:structfunc}a) and in the CMZ gas stream (\ref{Figure:structfunc}d) exhibit local minima at specific spatial scales. Structure functions display localised minima in response to periodicity in spatial or temporal data, indicating that the observed velocity fluctuations oscillate with an intrinsic wavelength (for an intuitive demonstration of this behaviour, see the Supplementary Information and \ref{fig:edf1}). We measure the wavelengths of these oscillations to be $405^{+92}_{-76}$\,pc and $22.0^{+5.4}_{-6.3}$\,pc in the spiral arm and CMZ gas stream, respectively. The coherent nature of these oscillatory motions, detected over 10$^{2}$-10$^{3}$\,pc scales, is inconsistent with the characteristically scale-free motion produced in a turbulent flow.  

Inspection of the corresponding density structure functions reveals periodicity on equivalent spatial scales to that detected in velocity. In the spiral arm, we find a minimum located at $366^{+88}_{-77}$\,pc. In the CMZ, two minima are identified at $6.0^{+0.8}_{-0.6}$\,pc and $21.8^{+5.5}_{-6.3}$\,pc, respectively. The periodicity implied by these localised minima corresponds to the regular spacing of molecular cloud complexes detected in CO ($2-1$) emission in the spiral arm of NGC\,4321 and GMCs detected in dust continuum emission in the CMZ gas stream, respectively (emission profiles extracted along the crests of these structures are shown in \ref{fig:edfdenvel}). The derived separation of the cloud complexes in the spiral arm of NGC\,4321 agrees with the 410\,pc spacing of embedded star clusters independently identified using mid-infrared observations\cite{elmegreen18}. Multiple minima detected in the density structure function of the CMZ gas stream suggests that GMCs separated by $6.0^{+0.8}_{-0.6}$\,pc are clustered together in groups separated by $21.8^{+5.5}_{-6.3}$\,pc. This is confirmed via inspection of the distribution of N$_{2}$H$^{+}$ ($1-0)$ emission from which our velocity information is derived (see Supplementary Information). 

The phase difference between the density and velocity fluctuations encodes the physical origins of the detected gas flows. In the spiral arm, we find that the velocities of the molecular cloud complexes are almost always blue shifted with respect to the rotational velocity of the galaxy at this location. The molecular cloud complexes, detected as peaks in the CO ($2-1$) emission, therefore reside where the velocity gradient is close to zero (\ref{fig:edfvdiff}a). Combining knowledge of NGC\,4321's inclination and rotation\cite{elmegreen18, lang19}, with the location of our selected arm with respect to the corotation radius of the spiral pattern (7.1-9.1\,kpc\cite{elmegreen89}), we conclude that the observed oscillation likely results from a combination of spiral streaming motions and radial gravitational inflow\cite{meidt13}. Conversely, in the CMZ, the locations of N$_{2}$H$^{+}$ ($1-0)$ emission peaks spatially correlate with extrema in the velocity gradient (see \ref{fig:edfvdiff}). This signature implies that the gas flows are converging towards the emission peaks. In this context, the GMCs observed in the CMZ gas stream may be the result of hierarchical collapse\cite{vazquez-semadeni19}, whereby large scale cloud complexes fed by flows on $22.0^{+5.4}_{-6.3}$\,pc scales have fragmented into a series of GMCs separated by $6.0^{+0.8}_{-0.6}$\,pc. Further discussion on the nature of these flows can be found in the Supplementary Information. 

As molecular clouds form at the stagnation points of convergent flows, the kinetic energy cascading from large to small scales contributes to the supersonically turbulent motions that produce complex and scale-free structure observed within GMCs\cite{klessen10, padoan14}. Indeed, in contrast to the characteristic scaling observed in the density and velocity structure described above, the internal physical structure of our selected GMCs in the Galactic disc and in the CMZ is multi-dimensional, complex, and without regularity (see \ref{fig:edfmaps}b and e). This is confirmed in \ref{Figure:structfunc}b and e, where we show that the structure functions of both density and velocity scale as power-laws. This scale-free behaviour is also reflected in structure functions computed over discrete azimuthal angles, indicating that the velocity fluctuations do not exhibit a preferred orientation (see Methods).

There is broad agreement between the scaling of our velocity structure functions (${0.41\pm0.01}$ and ${0.37\pm0.01}$ for the GMCs in the disc and CMZ, respectively) and the distribution of measured scaling exponents for the Galactic line width-size relationship\cite{heyer15}, which is often interpreted as evidence for the presence of supersonic turbulence in molecular clouds\cite{larson81}. However, we caution that the scaling of velocity structure functions has been interpreted in many different ways and may be subject to observational biases (see Supplementary Information). Despite the similarity in the scaling exponents of the two regions, the velocity fluctuations in the CMZ are greater than those in the disc by a factors of 10-100 when measured on a fixed spatial scale, consistent with the elevated levels of turbulence present within the inner Galaxy\cite{heyer15, henshaw16}. 

Shocks generated by supersonic turbulence contribute to the formation of the filamentary structure that pervades molecular clouds\cite{andre14, padoan14}. High-resolution observations of one of the density enhancements detected in our Galactic disc GMC, reveal a complex internal network of dense, velocity-coherent filaments\cite{henshaw14} (see \ref{fig:edfmaps}b and c). Embedded within each of these filaments is a population of density enhancements representing the formation sites of individual stars and stellar systems\cite{henshaw14, henshaw16b}. In analogy to the gas flows observed on much larger scales along the spiral arm of NGC\,4321 and in the CMZ gas stream, we find that the velocity fluctuations observed along the spine of our selected GMC filament exhibit periodicity on $0.28^{+0.06}_{-0.08}$\,pc scales (\ref{Figure:structfunc}c). The wavelength of these velocity oscillations is comparable to the separation of density enhancements along the filament ($0.32^{+0.01}_{-0.01}$\,pc). Furthermore, the locations of some of these density enhancements spatially correlate with extrema in the velocity gradient (\ref{fig:edfvdiff}c), indicative of either convergent motion or collapse-induced rotation\cite{misugi19}. 

A promising candidate driving the formation of periodic density fluctuations, and their correlated gas flows, is gravitational instabilities. We find that the separation of the periodic density enhancements observed in the spiral arm, CMZ gas stream, and GMC filament are factors of $3-5$ times the beam-deconvolved diameters of their parent structures ($122\,\pm\,5$\,pc, $4.2\,\pm\,0.2$\,pc, and $0.107\,\pm\,0.001$\,pc, respectively; see \ref{tab:lengthscales}). Periodic density enhancements arranged like `beads on a string' along their parent filaments have been observed in nearby galaxy discs\cite{elmegreen83,elmegreen18} as well as in local molecular clouds\cite{schneider79,tafalla15}, and frequently show separation-to-diameter ratios consistent with those measured in this study (see Supplementary Information). This measured ratio is consistent with theoretical work describing the gravitational fragmentation of filaments, both on large\cite{elmegreen94} and small\cite{nagasawa87, inutsuka92} scales. Our findings now extend this result by showing that the separation-to-diameter ratio in the velocity structure of the ISM matches that of the density enhancements. 

The combined analysis of density structure and gas kinematics, as well as the unprecedented spatial dynamic range covered in our study, represents a novel approach to understanding gas flows in the ISM and their relation to the emergence of physical structure. Our results indicate that the formation and evolution of dense star-forming gas across this sample of environments is controlled by nested, interdependent gas flows. The natural next steps are to apply this approach to an unbiased sample of regions, as well as to synthetic observations of numerical simulations, which together will provide critical insight into the role that the galactic environment plays in setting the scale and magnitude of the gas flows that seed star formation in galaxies.

\newpage

%%%%%%%%%%%%%%%%%%%%%%%%%%%%%%%%%%%%%%%%%%%%%%%%
%%%%%%%%%%%%%%%%%    FIGS.    %%%%%%%%%%%%%%%%%%
%%%%%%%%%%%%%%%%%%%%%%%%%%%%%%%%%%%%%%%%%%%%%%%%

\renewcommand{\thefigure}{Figure~\arabic{figure}}
\renewcommand{\thetable}{Table~\arabic{table}}

\captionsetup[figure]{labelformat=empty}% redefines the caption setup of the figures environment in the beamer class.
\captionsetup[table]{labelformat=empty}% redefines the caption setup of the figures environment in the beamer class.

\begin{figure*}
\begin{center}
\includegraphics[trim = 0mm 50mm 0mm 20mm, clip, width = 1.0\textwidth]{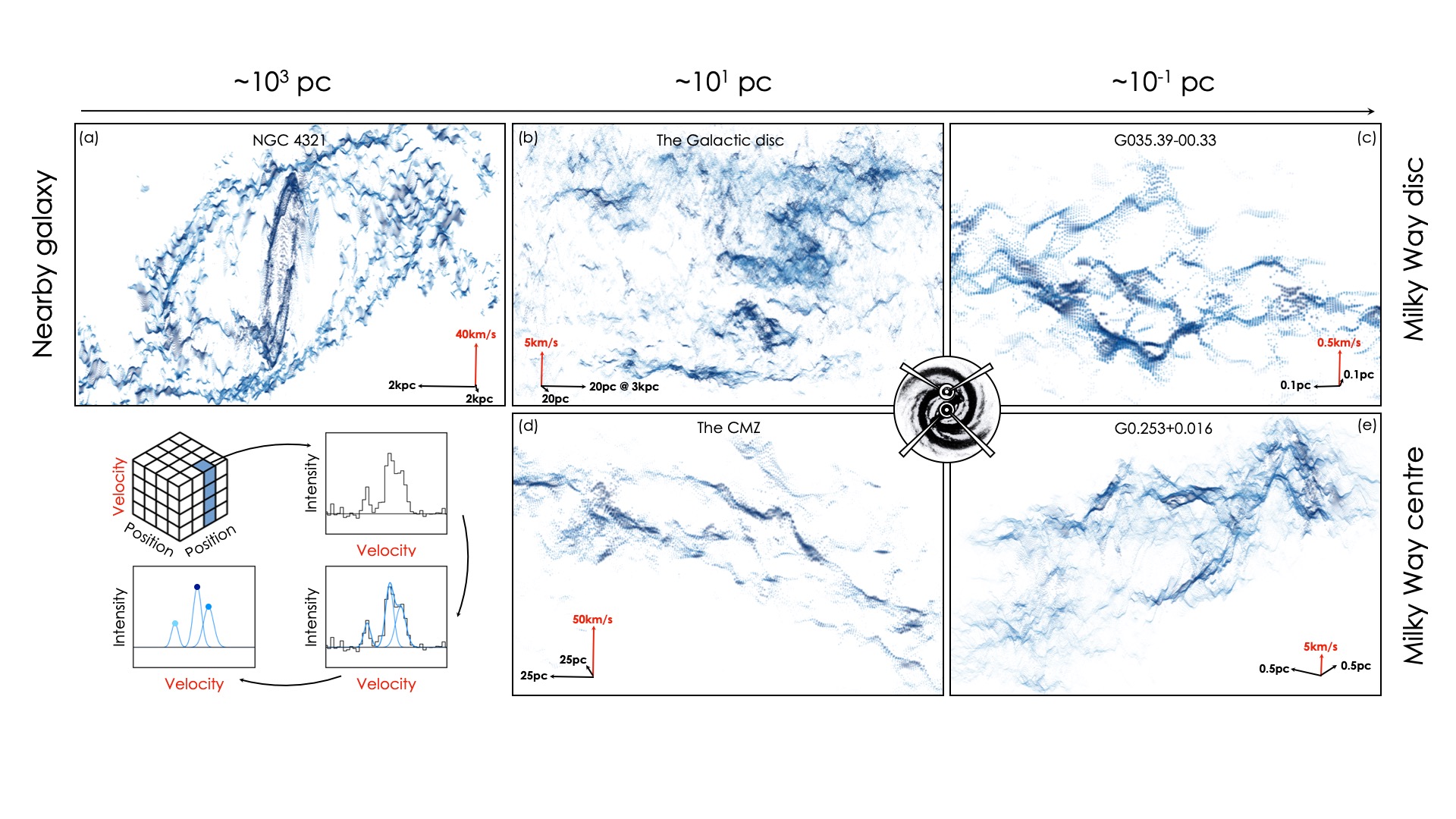}
%\vspace{-15mm}
\end{center}
\caption{\textbf{Figure 1: Ubiquitous velocity fluctuations throughout the molecular interstellar medium.} Here we show $p$-$p$-$v$ volumes of different galactic environments (see also the Supplementary Videos). The data points indicate the X-Y position and velocity of individual Gaussian emission features extracted from each data set using spectral decomposition (see Methods). The colour indicates the peak intensity of each emission feature (see the cartoon in the bottom-left). The different columns illustrate decreasing spatial scale, from kpc scales on the left down to sub-pc scales on the right. The rows highlight differences in galactic environment, from galaxy discs (top) to the Central Molecular Zone (CMZ; bottom). The individual regions in each panel are as follows: NGC\,4321 (a); a region in the Milky Way's Galactic disc (b); the Northern part of the infrared dark cloud (IRDC), G035.39--00.33 (c); the inner 250\,pc of the Milky Way, the CMZ (d); and the IRDC G0.253+0.016 (e). We include 3-D scale bars in the bottom corners of each panel to indicate the physical scaling as well as the orientation of each $p$-$p$-$v$ volume. Note that the our selected region in the Galactic disc (b) contains gas located at different distances. The scale bar is correct for a distance of 3\,kpc, which is relevant for the statistical analysis of our selected GMC (see Methods). }
\label{Figure:wiggles}
\vspace{-4mm}
\end{figure*}

\begin{figure*}
\begin{center}
\includegraphics[trim = 0mm 40mm 0mm 20mm, clip, width = 1.0\textwidth]{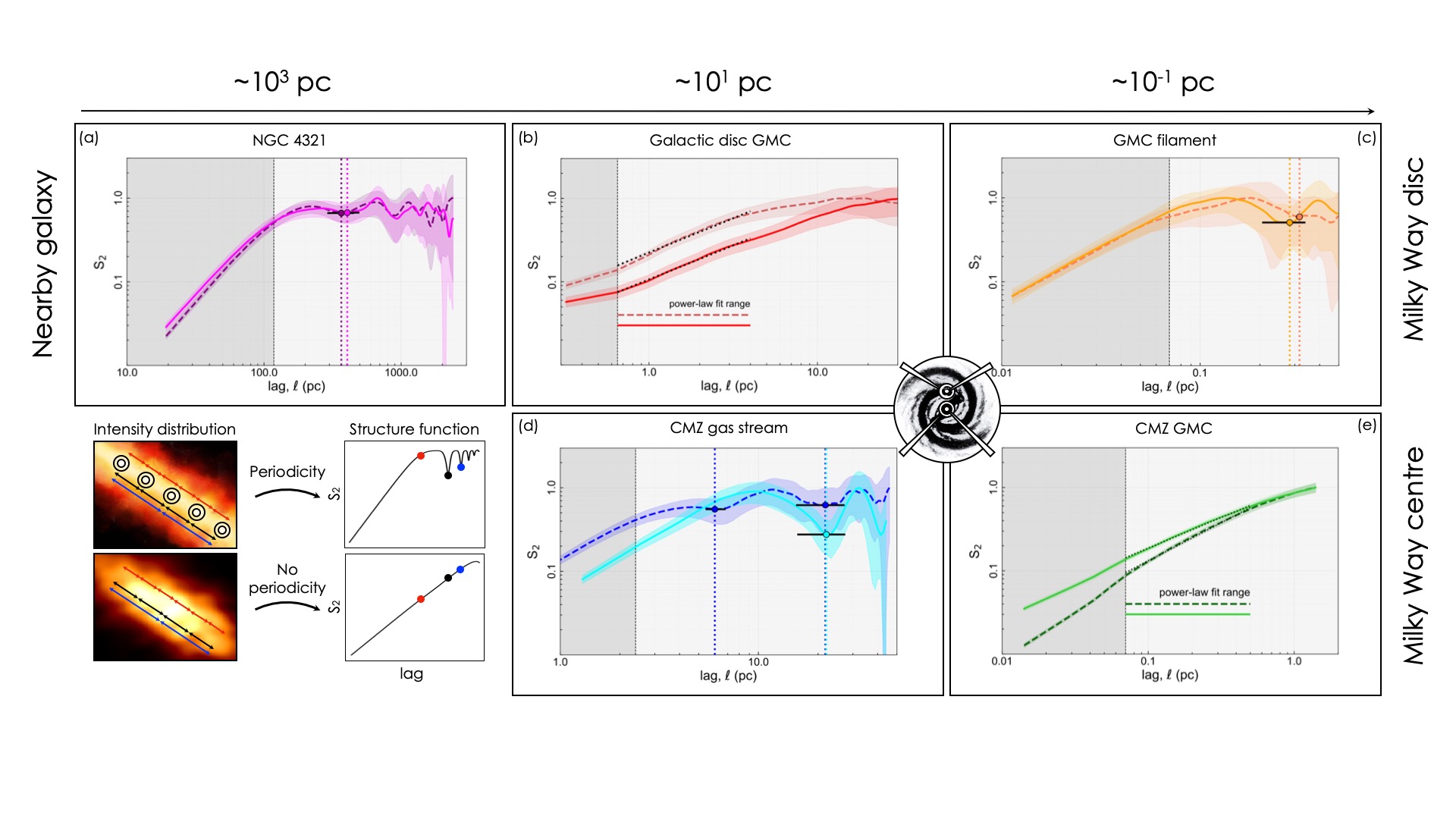}
\vspace{-10mm}
\end{center}
\caption{\textbf{Figure 2: Correlated density and velocity fluctuations.} Each panel shows normalised and noise-corrected second-order structure functions of density (coloured dashed lines) and velocity (coloured solid lines) taken from a sub-sample of the data sets presented in \ref{Figure:wiggles}. The regions selected for this analysis are shown in \ref{fig:edfmaps} (see Methods) and consist of: a portion of the southernmost dominant spiral arm in NGC\,4321 (a); a GMC located within the Galactic Disc (b); an individual filament located within the same GMC (c); a portion of the CMZ gas stream (d); a GMC located within the CMZ (e). Vertical coloured dotted lines in panels (a), (c), and (d) indicate the spatial scale of periodic fluctuations in density and velocity (the horizontal black lines indicate the uncertainty on these measurements; \ref{tab:lengthscales}). Black dotted lines in panels (b) and (e) are power-law fits to the structure functions (the fit range is indicated in each panel). The dark grey shaded area indicates the range of lags for which the recovered structure function is unreliable owing to the limited spatial resolution of the data.  }
\label{Figure:structfunc}
\vspace{-4mm}
\end{figure*}

\clearpage

\begin{table*}
 \centering
  \begin{minipage}{145mm}
  \begin{tabular}{cccc}
   \hline\\[-1.5ex]
   Environment & Filament Diameter & Density Periodicity & Velocity Periodicity  \\[1ex] 
   & (parsec) & (parsec) & (parsec)  \\[1ex] 
   \hline\\[-2ex]
   Spiral arm & $122\pm5$ & $366^{+88}_{-77}$ & $405^{+92}_{-76}$ \\[1ex] 
   CMZ Gas Stream & -- &  $6.0^{+0.8}_{-0.6}$ & --  \\[1ex] 
   & $4.2\pm0.2$ & $21.8^{+5.5}_{-6.3}$ & $22.0^{+5.4}_{-6.3}$ \\[1ex] 
   GMC Filament & $0.107\pm0.001$ & $0.32^{+0.01}_{-0.01}$ & $0.28^{+0.06}_{-0.08}$ \\[1ex] 
   \hline
  \end{tabular} 
  \end{minipage}
  \caption{\label{tab:lengthscales}\textbf{Table 1: Characteristic length scales.} In this table we include the beam-deconvolved diameters of the filamentary structures that exhibit periodicity, as well as the separation between periodically-spaced density enhancements and the wavelength of the velocity oscillations determined from our structure function analysis presented in \ref{Figure:structfunc}. Note that the density and velocity fluctuations observed throughout the GMCs selected in our study have no discernible characteristic scaling and are therefore not included here.}
\end{table*}  

%%%%%%%%%%%%%%%%%%%%%%%%%%%%%%%%%%%%%%%%%%%%%%%%%%%%%%%%%%%%%%%%%%%%%%%%%%%%%%%%%%%%%%%%%%%%%%%%
%%%%%%%%%%%%%%%%%%%%%%%%%%%%%%%%%%          Methods.           %%%%%%%%%%%%%%%%%%%%%%%%%%%%%%%%%
%%%%%%%%%%%%%%%%%%%%%%%%%%%%%%%%%%%%%%%%%%%%%%%%%%%%%%%%%%%%%%%%%%%%%%%%%%%%%%%%%%%%%%%%%%%%%%%%

\clearpage 
\noindent\HorRule\color{black}

\setcounter{figure}{0}
\setcounter{table}{0}
\captionsetup[figure]{labelformat=empty}% redefines the caption setup of the figures environment in the beamer class.
\captionsetup[table]{labelformat=empty}% redefines the caption setup of the figures environment in the beamer class.
\renewcommand{\thefigure}{Extended Data Figure~\arabic{figure}}
\renewcommand{\thetable}{Extended Data Table~\arabic{table}}

\begin{center}
{\bf \Large \uppercase{Methods} }
\end{center}

%%%%%%%%%%%%%%%%%%%%%%%%%%%%%%%%%%%%%%%%%%%%%%%%
%%%%%%%%%%%%%%%  ENVIRONMENT.   %%%%%%%%%%%%%%%%
%%%%%%%%%%%%%%%%%%%%%%%%%%%%%%%%%%%%%%%%%%%%%%%%

\section*{Environment selection}

Our observations cover a spatial dynamic range of about four orders of magnitude, from kpc scales in the nearby galaxy NGC\,4321 down to 0.1\,pc scales in the Milky Way. In the following section we describe our environment selection. 

\subsubsection*{NGC\,4321} 

Located in the Virgo cluster, NGC~4321 (M100) is a barred grand-design spiral galaxy of class SAB(s)bc\cite{de-vaucouleurs91}. The galaxy is highly structured with two well-defined spiral arms with strong symmetry\cite{elmegreen11}. The galaxy has a stellar bar and a nuclear ring with a radius of about 1~kpc\cite{knapen95}. Strings of regularly-spaced star-forming regions extend over kpc distances within thin dust filaments throughout its disc\cite{knapen96, elmegreen18}. 

\subsubsection*{The Galactic disc} 

On intermediate scales in the Milky Way disc we include observations of the massive ($5\times10^{5}$\,\msun; Ref.~\citenum{ragan14}) Giant Molecular Filament, GMF\,38.1--32.4b\cite{ragan14}. GMFs are a class of elongated giant molecular clouds\cite{goodman14, ragan14, zucker15, abreu-vicente16}. Our selection of GMF\,38.1--32.4b is based on its association with the Infrared Dark Cloud (IRDC) G035.39--00.33, which provides the basis for the high-resolution component of our study of galaxy discs (see below). GMF\,38.1--32.4b is almost orthogonal to the Galactic plane and has a length of 80\,pc and an aspect ratio about 1:12\cite{ragan14,zucker18}. 

\subsubsection*{G035.39--00.33} 

Situated at a kinematic distance of 2.9\,kpc\cite{simon06}, G035.39--00.33 is a massive (10$^{4}$\,\msun\cite{kainulainen13}) and filamentary IRDC embedded within GMF\,38.1--32.4b and thought to harbour the early stages of star formation\cite{jimenez-serra10, nguyen-luong11}. The structure and dynamics of G035.39--00.33 have been studied extensively, revealing the presence of multiple sub-filaments which feed their embedded core population\cite{henshaw13, henshaw14, henshaw16b, henshaw17, jimenez-serra14, sokolov19}. 

\subsubsection*{Central Molecular Zone (CMZ)} 

The inner few $100~\pc$ of the Milky Way contains approximately 3-5\% of the Milky Way's molecular gas, a reservoir of $2-7\times10^{7}$\,\msun \ of molecular material\cite{ferriere07, heyer15}. The physical conditions of the gas are extreme compared to those in the galaxy discs discussed above: the density\cite{longmore12, longmore13, mills18}, temperature\cite{ginsburg16, krieger17}, velocity dispersion\cite{shetty12, henshaw16}, radiation field\cite{clark13}, pressure\cite{kruijssen13, walker18} and cosmic ray ionization rate\cite{yusef-zadeh07} are larger by factors of a few to several orders of magnitude. The gas is distributed throughout several coherent streams spanning 250\,\kms \ in velocity and with projected lengths of the order 100-250\,pc\cite{sofue95, molinari11, kruijssen15, henshaw16}. 

\subsubsection*{G0.253+0.016} 

On small scales in the CMZ, we focus on the IRDC G0.253+0.016. With a mass of around $10^{5}$\,M$_{\odot}$ and an equivalent radius of just $\sim2-3$\,pc, G0.253+0.016 is one of the densest molecular clouds in the Galaxy\cite{longmore12, kauffmann13, rathborne15}. Despite this, G0.253+0.016 shows very few signatures of active star formation\cite{mills15}. While single dish observations depict G0.253+0.016 as a single, coherent, and centrally-condensed molecular cloud\cite{rathborne14}, more recent work suggests that G0.253+0.016 is dynamically complex and hierarchically-structured\cite{henshaw19}. 

%%%%%%%%%%%%%%%%%%%%%%%%%%%%%%%%%%%%%%%%%%%%%%%%
%%%%%%%%%%%%%%%%%     DATA.   %%%%%%%%%%%%%%%%%%
%%%%%%%%%%%%%%%%%%%%%%%%%%%%%%%%%%%%%%%%%%%%%%%%

\section*{Observations and data} 

\begin{figure*}
\begin{center}
\includegraphics[trim = 0mm 0mm 0mm 0mm, clip, width = 0.98\textwidth]{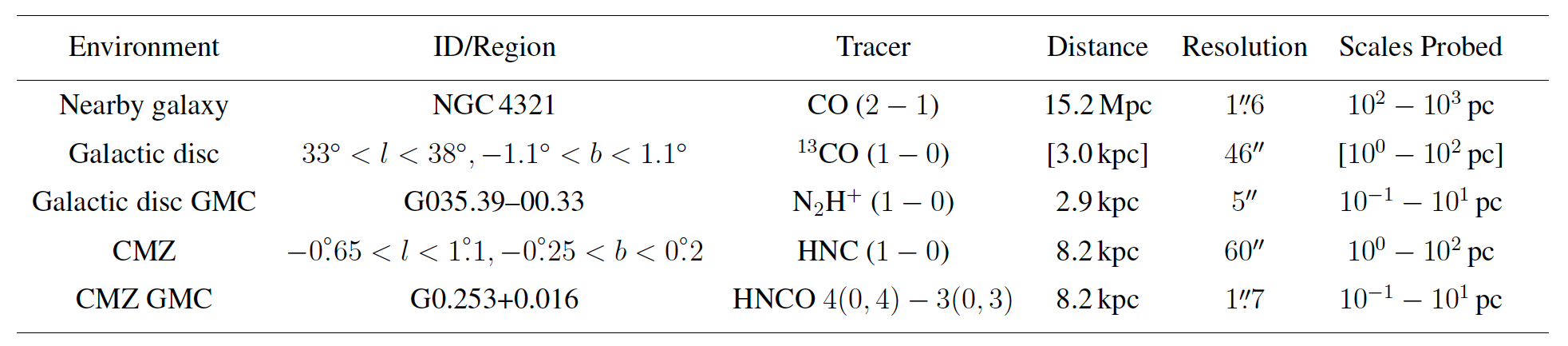}
\end{center}
\vspace{-2mm}
\caption{\textbf{Extended Data Figure 1 $|$ Summary of the observations.} Here we highlight the observations and region selection for the data presented in \ref{Figure:wiggles}. The scales probed by our Galactic disc selection (seen in square brackets) are relevant for a distance of 3\,kpc. Out of each of these environments we select sub-regions for the statistical analysis presented in \ref{Figure:structfunc} (see {\bf Statistical analysis of the observational data}).}
\vspace{-4mm}
\label{Table:obs}
\end{figure*}

Here we describe the observations and data, summarising the relevant information in \ref{Table:obs}. 

\subsubsection*{NGC\,4321} 

We analyse CO~(2-1) emission from NGC~4321 (M100) observed by ALMA as part of the pilot program for the PHANGS-ALMA survey (A.~K.~Leroy et al., manuscript in preparation). These Cycle 3 (program 2015.1.00956.S) observations covered the galaxy using ALMA's main array of 12-m telescopes, the 7-m telescopes from the Morita Atacama Compact Array, and the total power dishes. The galaxy was covered by two large mosaics, which were observed separately. Data calibration used the standard ALMA pipeline. For imaging, we combined the data from the 12m and 7m arrays, carried out a multiscale deconvolution, and then ``feathered'' the interferometric images with the total power data to produce the final images. A first version of these maps initially appeared in Ref.~\citenum{sun18}. Sample selection, observing strategy, data reduction, imaging, and post-processing are described in A.~K.~Leroy et al., (manuscript in preparation). 

Relevant to this work, these data include total power and short spacing information and so have sensitivity to all spatial scales. For imaging, we binned the data to have a channel width of $2.5$~km~s$^{-1}$. During post-processing, two mosaics were convolved to share a matched $\sim 1\farc6$ beam ($\sim 120$~pc at the adopted distance to NGC~4321 of 15.2\,Mpc\cite{tully09}) and then linearly combined to form a single data cube covering most bright CO emission from the galaxy. Following Ref.~\citenum{sun18}, $\sim 70\%$ of the total CO emission present in the target region is recovered at good signal to noise at $1\farc6$ resolution.

\subsubsection*{The Galactic disc}

We use data from the Boston University Five College Radio Astronomy Observatory (FCRAO) Galactic Ring Survey (GRS\cite{jackson06}). This survey covered the lowest rotational transition of the $^{13}$CO isotopologue ($J=1\rightarrow0$) with an angular resolution of $46^{\prime\prime}$, a pixel sampling of $22^{\prime\prime}$, and a spectral resolution of $0.21$~\kms. At a distance of 3\,kpc, relevant for GMF\,38.1--32.4b\cite{ragan14}, the angular resolution of these data corresponds to a physical resolution of about 0.7\,pc. The GRS covers a range in Galactic coordinates of $14\fdg0<l<56\fdg0$ and $-1\fdg 1<b<1\fdg 1$. The region displayed in the upper central panel of \ref{Figure:wiggles} thereby encloses both GMF\,38.1--32.4b and G035.39--00.33 and covers $33\fdg0<l<38\fdg0$, the full Galactic latitude coverage, and $25 \leq v_{\text{LSR}} \leq 70$ \kms. Note that because of our view through the Galactic plane this panel contains molecular gas situated at different distances, and GMF\,38.1--32.4b comprises only a small fraction of these data (see {\bf Data selection for statistical analysis}).

\subsubsection*{G035.39--00.33} 

We analyse N$_{2}$H$^{+}$ ($1-0$) emission associated with G035.39--00.33 observed with the Institut de Radioastronomie Millim\'{e}trique (IRAM) Plateau de Bure Interferometer. These data were first presented in Refs.~\citenum{henshaw14} and \citenum{henshaw16b}. The observations cover the northern portion of the cloud and have an angular extent of $40^{\prime\prime} \times 150^{\prime\prime}$ ($0.6\,{\mathrm{pc}}\times2.1\,{\mathrm{pc}}$; assuming a kinematic distance of 2.9\,kpc\cite{simon06}). These data were combined with IRAM-30m telescope observations first presented in Ref.~\citenum{henshaw13} to recover the zero-spacing information. The data are convolved to a circular beam of $5^{\prime\prime}$ (the native spatial resolution is $3\overset{\prime\prime}{.}9 \times3\overset{\prime\prime}{.}2$), corresponding to a physical resolution of 0.07\,pc. In this study, we retain the native pixel scaling of $0\overset{\prime\prime}{.}76$ (Ref.~\citenum{henshaw14} downsampled the data to $2^{\prime\prime}$ pixels). The spectral resolution of the data is 0.14\,km\,s$^{-1}$, and the sensitivity is of the order 1\,mJy\,beam$^{-1}$. Further details on these observations and their combination are included in Ref.~\citenum{henshaw14}. 

\subsubsection*{Central Molecular Zone (CMZ)} 

We use data from the Mopra CMZ survey\cite{jones12}. These observations cover the region $-0\fdg65<l<1\fdg1$ and $-0\fdg25<b<0\fdg20$, which incorporates material within a galactocentric radius of approximately 125\,pc (assuming a distance of 8340\,pc\cite{reid14}). The spatial resolution of the observations is $60^{\prime\prime}$, which corresponds to a physical resolution of 2.4\,pc. The spectral resolution of the observations is 2\,\kms. The data included in \ref{Figure:wiggles}d show the HNC $(1-0)$ emission analysed in Ref.~\citenum{henshaw16}, which is extended over the entire CMZ. However, note that for our statistical study we use N$_{2}$H$^{+}$ ($1-0$), consistent with Ref.~\citenum{henshaw16c} (see {\bf Spectral decomposition}). We refer the reader to Refs.~\citenum{henshaw16} and \citenum{jones12} for further discussion on these data.

The \emph{Herschel} column density data was computed using modified blackbody fits to the HiGAL data\cite{molinari10}, using the 160, 250, 350, and 500~$\mu$m~bands (C.~Battersby et al., manuscript in preparation). The 70~$\mu$m~band was excluded from the fit due to possible contamination from very small, warm dust grains. At the wavelengths observed with Herschel, there is significant contamination from Galactic cirrus emission, which was removed through an iterative process described in Refs.~\citenum{battersby11} and \citenum{mills17}. All of the data were smoothed to the HiGAL reported beam size at the longest wavelength, 36$^{\prime\prime}$ (Ref.~\citenum{molinari10}), and the resulting 
maps are presented at this resolution. 

\subsubsection*{G0.253+0.016} 

We include the ALMA Cycle~0 observations (program 2011.0.00217.S) of IRDC G0.253+0.016 studied in detail by Ref.~\citenum{henshaw19}. These data were first presented in Ref.~\citenum{rathborne15}. The observations cover the full $180^{\prime\prime} \times 60^{\prime\prime}$ ($7.1{\mathrm{pc}}\times2.4\,{\mathrm{pc}}$; assuming a distance of 8340\,pc\cite{reid14}) extent of the molecular cloud. We focus exclusively on the HNCO $4(0,4)-3(0,3)$ transition. These data were combined with single dish observations from the The Millimeter Astronomy Legacy Team 90 GHz (MALT90) survey\cite{foster11,jackson13} to recover the zero-spacing information. The spatial resolution of the data is $1\farc7$, which corresponds to a physical resolution of $0.07$\,pc. The spectral resolution of the data is 3.4\,\kms. 

%%%%%%%%%%%%%%%%%%%%%%%%%%%%%%%%%%%%%%%%%%%%%%%%
%%%%%%%%%%%%%%   DECOMPOSITION.  %%%%%%%%%%%%%%%
%%%%%%%%%%%%%%%%%%%%%%%%%%%%%%%%%%%%%%%%%%%%%%%%

\section*{Spectral decomposition}

In this section we describe the spectral decomposition of the data discussed above. For an introduction to the methodology and a description of the techniques used, we refer the reader to the Supplementary Information. The spectral decomposition of the CMZ fields displayed in \ref{Figure:wiggles}d and e is described in full in Refs.~\citenum{henshaw16} and \citenum{henshaw19}, respectively. Although \ref{Figure:wiggles}d shows the HNC $(1-0)$ decomposition to better highlight extended emission, we use the decomposition of N$_{2}$H$^{+}$ $(1-0)$ emission for the structure function analysis presented in \ref{Figure:structfunc}d, consistent with Ref.~\citenum{henshaw16c}. 

New to this work, we apply {\sc ScousePy} to the CO $(2-1)$ data towards NGC\,4321. We set the width of our SAAs to 100 pixels, corresponding to about 1.8\,kpc at the assumed distance of NGC\,4321 (15.2\,Mpc). A total of $1.1\times10^{5}$ model solutions were obtained out of the $2.3\times10^{5}$ included in the SAA coverage. A total of $1.2\times10^{5}$ Gaussian components were extracted during the fitting procedure, indicating that the models were mostly single-component fits. Multiple component fits were largely confined to the inner part of the galaxy and a few concentrated regions in the spiral arms. 

The decomposition of the Galactic disc region, focussing on GMF\,38.1--32.4b\cite{ragan14}, is performed using the machine learning algorithm {\sc GaussPy+}\cite{riener19} (see Supplementary Information). The details of the decomposition of the entire $^{13}$CO ($1-0$) GRS data set are described in full in Ref.~\citenum{riener20}. We train {\sc GaussPy} with twelve training sets each containing 500 randomly chosen spectra from the GRS data set. These training sets are automatically generated with {\sc GaussPy+}, and their decomposition was benchmarked against the training set functionality of {\sc ScousePy}\cite{henshaw19}, which takes randomly sampled regions of survey data for training set development. We find good agreement between the results of the two methods in these sub-regions, justifying our application of {\sc GaussPy+}. \ref{Figure:wiggles}b shows a subsample of the GRS decomposition containing almost $3\times10^{5}$ components within $33^{\circ}<l<38^{\circ}$, the full Galactic latitude coverage, and $25 \leq v_{\text{LSR}} \leq 70$ \kms. 

For G035.39--00.33, the spectral decomposition is performed using {\sc ScousePy} and therefore differs from that originally presented in Ref.~\citenum{henshaw14}. We perform our Gaussian decomposition focusing exclusively on the isolated $F_{1}, F = 0,1 \rightarrow 1,2$, component of the $J=1\rightarrow0$ transition of N$_{2}$H$^{+}$ at $93176.2522$\,MHz\cite{pagani09}. We set the width of our SAAs to 16 pixels, corresponding to about 0.2\,pc at the assumed distance of G035.39--00.33 (2.9\,kpc). A total of $1.5\times10^{4}$ velocity components were fitted to $7\times10^{3}$ pixels, and multiple velocity components are required to describe the spectral line profiles over a significant (69\%) portion of the map. 

%%%%%%%%%%%%%%%%%%%%%%%%%%%%%%%%%%%%%%%%%%%%%%%%
%%%%%%%%%%%%%%  DATA SELECTION.  %%%%%%%%%%%%%%%
%%%%%%%%%%%%%%%%%%%%%%%%%%%%%%%%%%%%%%%%%%%%%%%%

\section*{Data selection for statistical analysis}

In the following section we describe our data selection for the statistical analysis presented in \ref{Figure:structfunc}. A potential source of uncertainty in interpreting the results from kinematic analysis methods is that observational data is sensitive only to estimators of the true underlying density and velocity fields. Spectral decomposition, which relies on the profile of emission lines, can be vulnerable to the influence of velocity crowding\cite{burton72, lazarian00, ostriker01} and variations in optical depth. Fortunately, the influence of these effects can be mitigated through careful selection of environment and spectral lines. 

We select sub-regions from each of the five environments discussed in {\bf Environment selection} and presented in \ref{Figure:wiggles}. We select three sub-regions which, despite tracing vastly different scales, display similar morphology in that they are highly filamentary and (qualitatively) interspersed with quasi-periodic intensity peaks along their crests: part of the southern spiral arm in NGC\,4321, part of the CMZ gas stream, and a filament embedded within a GMC located in the Galactic disc. We further select two GMCs, one in the Galactic disc and one in the Milky Way's CMZ, the intensity profiles of which are complex and disordered. Our selected regions are displayed in \ref{fig:edfmaps}.

\subsubsection*{Spiral arm} 

For our analysis of NGC\,4321, we select a region which shows a sample of regularly-spaced infrared peaks\cite{elmegreen18}. These star-forming complexes, inferred by the presence of mid-infrared emission, are located within NGC\,4321's southern spiral arm (see green X's in Figure\,4 of Ref.~\citenum{elmegreen18}). 

We expect that the influence of velocity crowding is minimised in this region by the limited molecular scale height of the galaxy, which is viewed face on. Similarly, although the CO $(2-1)$ is most likely optically thick, single velocity components provide a suitable description for most of the CO $(2-1)$ emission along the arm. This indicates that our centroid measurements are not strongly influenced by optical depth effects. Where two velocity components are necessary to model the emission, the secondary component, i.e. that which appears in addition to the component most closely tracing the spiral arm, is often spatially localised and compact, of low brightness temperature, and offset in velocity from the emission tracing the arm. We remove these additional components by selecting the brightest velocity component at each location. We further select the molecular cloud complexes which follow the coherent structure of the spiral using {\sc Glue}\cite{glue15, glue17}. \ref{fig:edfmaps}a shows the data we have selected for our statistical analysis. 

\subsubsection*{Galactic disc GMC} 

The effect of velocity crowding is particularly pertinent in studies of the Milky Way, where our view through the Galactic plane complicates our physical interpretation of \ppv-space. Molecular clouds, coherent in true 3-D space may be separated in \ppv-space due to complex dynamics. Similarly, molecular clouds projected along the same line-of-sight, but otherwise physically decoupled, may crowd in \ppv-space. These effects are expected to be most prominent towards the tangent point velocities of the Milky Way's rotation curve\cite{riener20}. Since GMF\,38.1--32.4b is separated by around $40$\,\kms \ from the tangent point velocity at $l=35^{\circ}$,\cite{reid19} we expect the effects of velocity crowding to be small.

The gas associated with GMF\,38.1--32.4b makes up only a small fraction of the total emission displayed in \ref{Figure:wiggles}b. To isolate this GMC from the sample of data shown in \ref{Figure:wiggles}b, we use {\sc acorns}\cite{henshaw19}, an n-dimensional unsupervised clustering algorithm designed for the analysis of spectroscopic $p$-$p$-$v$ data\cite{henshaw19}. We apply {\sc acorns} directly to the data shown in \ref{Figure:wiggles}b. We cluster the data based on position, centroid velocity, and the velocity dispersion information provided by our decomposition. From the resulting hierarchy of clusters, we identify the cluster which most closely matches the morphology of GMF\,38.1--32.4b as identified by Ref.~\citenum{ragan14}. The result of this analysis is displayed in \ref{fig:edfmaps}b, where we show the peak brightness temperature of the $^{13}$CO $(1-0)$ emission of the {\sc acorns} cluster. Our {\sc acorns}-identified cluster shows excellent agreement with the GMF extracted in Ref.~\citenum{ragan14}, which is highlighted by the white contour. 

\subsubsection*{GMC filament} 

The association between GMF\,38.1--32.4b and IRDC G035.39--00.33 is evident in \ref{fig:edfmaps}b, where G035.39--00.33 is identified as a compact, bright source of $^{13}$CO $(1-0)$ emission. We apply {\sc acorns} to our new decomposition of the N$_{2}$H$^{+}$ $(1-0)$ data. Our clustering analysis is consistent with the results of Ref.~\citenum{henshaw14} in that the N$_{2}$H$^{+}$ $(1-0)$ emission is mainly distributed throughout three dominant sub-filaments (see also Ref.~\citenum{sokolov19}). This high-density gas tracer is less susceptible to the effects of velocity crowding and line-of-sight confusion\cite{clarke18}. Furthermore, the isolated hyperfine component of N$_{2}$H$^{+}$ $(1-0)$ is measured to be optically thin\cite{henshaw14}. The sub-filament selected for our structure function analysis is shown in \ref{fig:edfmaps}c. The filament exhibits strong localised velocity fluctuations along its primary axis (evident in \ref{Figure:wiggles}c).

\subsubsection*{CMZ} 

As proof of concept of our statistical analysis, we reanalyse the region of the CMZ studied by Ref.~\citenum{henshaw16c}. In general, the effects of velocity crowding are less prominent in the CMZ, where the gas is distributed throughout molecular streams that are well-separated in velocity\cite{kruijssen15, henshaw16}. Our selected region is located between $-0\fdg65<l<0\fdg0$ and $-0\fdg05<b<0\fdg1$, and is associated both with a series of quasi-regularly spaced molecular cloud condensations and an oscillatory pattern in the centroid velocity of the molecular gas. This velocity pattern is observed in multiple tracers\cite{henshaw16, henshaw16c,langer17, longmore17}, indicating that it is of dynamical origin, and not simply the result of excitation or optical depth effects. \ref{fig:edfmaps}d displays the \emph{Herschel}-derived column density map (C.~Battersby et al., manuscript in preparation) covering the region that displays the coherent velocity oscillation investigated in Ref.~\citenum{henshaw16c}. These data were selected using the software {\sc Glue}\cite{glue15, glue17}, using the location of the centroid velocity measurements from Ref.~\citenum{henshaw16c} as a guide for masking. 

\subsubsection*{CMZ GMC} 

We focus our analysis on one of G0.253+0.016's dominant substructures, extracted using {\sc acorns}\cite{henshaw19}. The distribution of optically-thin\cite{henshaw19} emission from the 4(0,4)--3(0,3) transition of HNCO throughout our selected substructure, labelled `tree C' in Ref.~\citenum{henshaw19}, is morphologically similar to that of the dust continuum emission associated with the cloud. This structure therefore most likely dominates the internal physical composition of G0.253+0.016. The HNCO $4(0,4)-3(0,3)$ emission profile of this substructure is displayed in the bottom-right panel of \ref{fig:edfmaps}e. 

\section*{Statistical analysis of the observational data} 

In the following section we discuss our statistical analysis of our selected sub-regions. We group the sub-regions according to their dominant geometry, either long and filamentary or multi-dimensional and complex. The physical interpretation of the following analysis is discussed at length in the section {\bf Summary of the results and physical interpretation} in the Supplementary Information. 

\subsubsection*{Analysis of filamentary structures} 

We perform our analysis along the crests of each sub-region. We obtain the crest of each filamentary structure by applying {\sc FilFinder}\cite{koch15} to our selected density tracer in each environment, CO ($2-1$) in the spiral arm, the \emph{Herschel}-derived column density map in the CMZ, and the N$_{2}$H$^{+}$ ($1-0$) emission in the GMC filament (see \ref{fig:edfmaps}). 

We perform a weighted mean of our selected density tracer (weighting by the square of the Gaussian fit amplitude value) orthogonal to the crest. The resulting profiles (solid) and the standard deviation about the mean (shaded region) are shown in the top panels of \ref{fig:edfdenvel}.

For the velocity field, we remove the bulk motion motion by modelling it with a simple polynomial function and subtracting this model from the velocity field derived from our spectral decomposition, leaving only the residual local velocity fluctuations. We measure a weighted mean velocity (weighting by the square of the Gaussian fit amplitude value) orthogonal to the crest. The resulting distributions (solid line) and the standard deviation about the mean (shaded region) are shown in the bottom panels of \ref{fig:edfdenvel}.

We compute the 1-D structure function on the data presented in \ref{fig:edfdenvel} using 
\begin{equation}
    S_{p}(\Bell)=\langle\delta x(\mathbf{r},\mathbf{\Bell})^{p}\rangle = \langle |x(\mathbf{r})-x(\mathbf{r}+\Bell)|^{p}\rangle,
\label{eq:sf}
\end{equation}
where the quantity $\delta x(\mathbf{r},\mathbf{\Bell})$ represents the absolute difference in the quantity $x$ measured between two locations separated by $\Bell$. The structure function, $S_{p}(\Bell)$, averages this quantity (raised to the $p$th power, the order) over all locations (indicated by the angle brackets). For a more comprehensive description of the structure function and its behaviour we refer the reader to {\bf Structure functions and their application to toy models} in the Supplementary Information. 

Since the structure function compares pairs of points at a given lag, $\Bell$, the maximum, fully sampled lag that fits within these 1-D data sets is half the total length of each crest. This defines our upper limit to the spatial scales over which the structure functions presented in \ref{Figure:structfunc} is computed. We generate noise-corrected structure functions by computing the structure function of the measurement uncertainties associated with our Gaussian fit components and subtracting this from the structure functions of the signal. A comprehensive description of the behaviour of the structure function in response to instrumental noise is included in the Supplementary Information. The results of this analysis are shown in \ref{Figure:structfunc}.

We estimate the uncertainty on the structure function as
\begin{equation}
\sigma_{S_{2}}(\Bell)=\frac{\sigma(\Bell)}{\sqrt{N_{\rm indep}}}
\end{equation}
where $\sigma$ represents the standard deviation of the measurements obtained at a given lag and $N_{\rm indep}$ is the number of independent measurements of $S_{2}$ taken at that same lag, $\Bell$.

Each of the structure functions relating to our filamentary structures display local minima at specific spatial scales. For our density tracers (dashed lines), these minima occur at spatial scales of $\lambda_{\rho}=366^{+88}_{-77}$\,pc in the spiral arm, both $6.0^{+0.8}_{-0.6}$\,pc and $21.8^{+5.5}_{-6.3}$\,pc in the CMZ, and $0.32^{+0.01}_{-0.01}$\,pc in the GMC filament (\ref{tab:lengthscales}). The uncertainties on each of these measurements represent the full-width-at-half-\emph{minimum} for each of the minima detected in the structure function. Periodicity in the density structure indicates that density enhancements are forming with a preferred, or characteristic, spacing. The structure function therefore provides a quantitative measure of the periodicity that is qualitatively evident in the maps presented in \ref{fig:edfmaps}. We demonstrate the response of the structure function to periodicity using toy models in the Supplementary Information (see e.g. \ref{fig:edf1}). 

We compute the beam-deconvolved diameters of the filamentary structures for comparison with the above derived characteristic spacing. To do this, we use FilFinder\cite{koch15} to compute radial intensity profiles normal to the spines identified above. We fit the radial profile with a Gaussian model with a mean centred on the filament spine and a constant background. The model is then deconvolved with the beam of the observations. For the spiral arm, CMZ stream, and GMC filament we measure beam-deconvolved diameters of $D=122\,\pm\,5$\,pc, $4.2\,\pm\,0.2$\,pc, and $0.107\,\pm\,0.001$\,pc, respectively. The characteristic wavelengths derived above are therefore of the order $3$, $5$, and $3$ times the diameter of their parent structures, respectively. 

Perhaps a more surprising feature of structure functions presented in \ref{Figure:structfunc} is that the velocity fluctuations along the crests of our selected regions also display periodic behaviour. Moreover, that the characteristic wavelengths of the velocity fluctuations in each environment agrees, within the uncertainties, to the periodicity detected in density. We measure the location of the minima in the velocity structure functions (solid lines) to be $\lambda_{v}=405^{+92}_{-76}$\,pc for the spiral arm, $22.0^{+5.4}_{-6.3}$\,pc for the CMZ, and $0.28^{+0.06}_{-0.08}$\,pc in the GMC filament (\ref{tab:lengthscales}). 

To assess the phase relationship between density ($\rho$) and velocity ($v$) fluctuations we introduce the cross-correlation, which is defined
\begin{equation}
XC(\Bell) = \langle \rho(\mathbf{r}) v(\mathbf{r}+\Bell)\rangle.
\label{eq:cc}
\end{equation}
The cross-correlation tells us how closely related the density and velocity fields are as a function of displacement, or lag, relative to one another. 

We define the associated uncertainty in the cross-correlation function as 
\begin{equation}
\begin{split}
%\sigma(Q)^2 = \sigma_{\rho(i)}^2 v(i+j)^2 + \sigma_{v(i+j)}^2 \rho(i)^2\\
\sigma_{XC}^2(\Bell) & = \Sigma_{\mathbf{r}}\left[\sigma_{\rho(\mathbf{r})}^2 v(\mathbf{r}+\Bell)^2\right] + \Sigma_{\mathbf{r}}\left[\sigma_{v(\mathbf{r}+\Bell)}^2 \rho(\mathbf{r})^2\right]
\end{split}
\end{equation}
where $\sigma_{\rho}$ and $\sigma_{v}$ represent the standard deviation of our measurements (shown as the coloured shaded regions in \ref{fig:edfdenvel}). Given that both variables (density and velocity) exhibit periodicity in each of our sub-regions, and are therefore autocorrelated, the cross-correlation function has multiple peaks at different lags. We therefore identify all significant ($XC(\Bell)/\sigma_{XC}(\Bell))>3$) peaks in the cross-correlation function, and select the one which is located at the smallest lag as the representative phase-shift between density and velocity.

We measure phase differences between density and velocity of $191\pm62$\,pc for the spiral arm, $2.7\pm2.9$\,pc for the CMZ, and $\-0.11\pm0.03$\,pc for the GMC filament, respectively. The uncertainty in each of these measurements represents the standard deviation of a Gaussian fitted to the relevant peaks in the cross-correlation function. 

Given that the spatial resolution of our density (column density from \emph{Herschel}) and velocity (N$_{2}$H$^{+}$ ($1-0$) emission from Mopra) tracers in the CMZ differ by a factor of two, we also compute the phase difference between density peaks and the velocity oscillations both derived from the N$_{2}$H$^{+}$ ($1-0$) emission and find $1.4\pm2.9$\,pc (the emission profile of the N$_{2}$H$^{+}$ data is displayed in Extended~Data~Figure~4b). Relative to the characteristic wavelengths derived for the velocity oscillations, $\lambda_{v}$, these phase differences are approximately $\lambda_{v}/2$ for the spiral arm, $\lambda_{v}/8$ ($\lambda_{v}/16$, for the N$_{2}$H$^{+}$ ($1-0$) emission) for the CMZ, and $2\lambda_{v}/5$, for the GMC filament, respectively. 

\subsubsection*{Analysis of multi-dimensional structures} 
\begin{figure*}
\begin{center}
\includegraphics[trim = 0mm 0mm 0mm 0mm, clip, width = 0.98\textwidth]{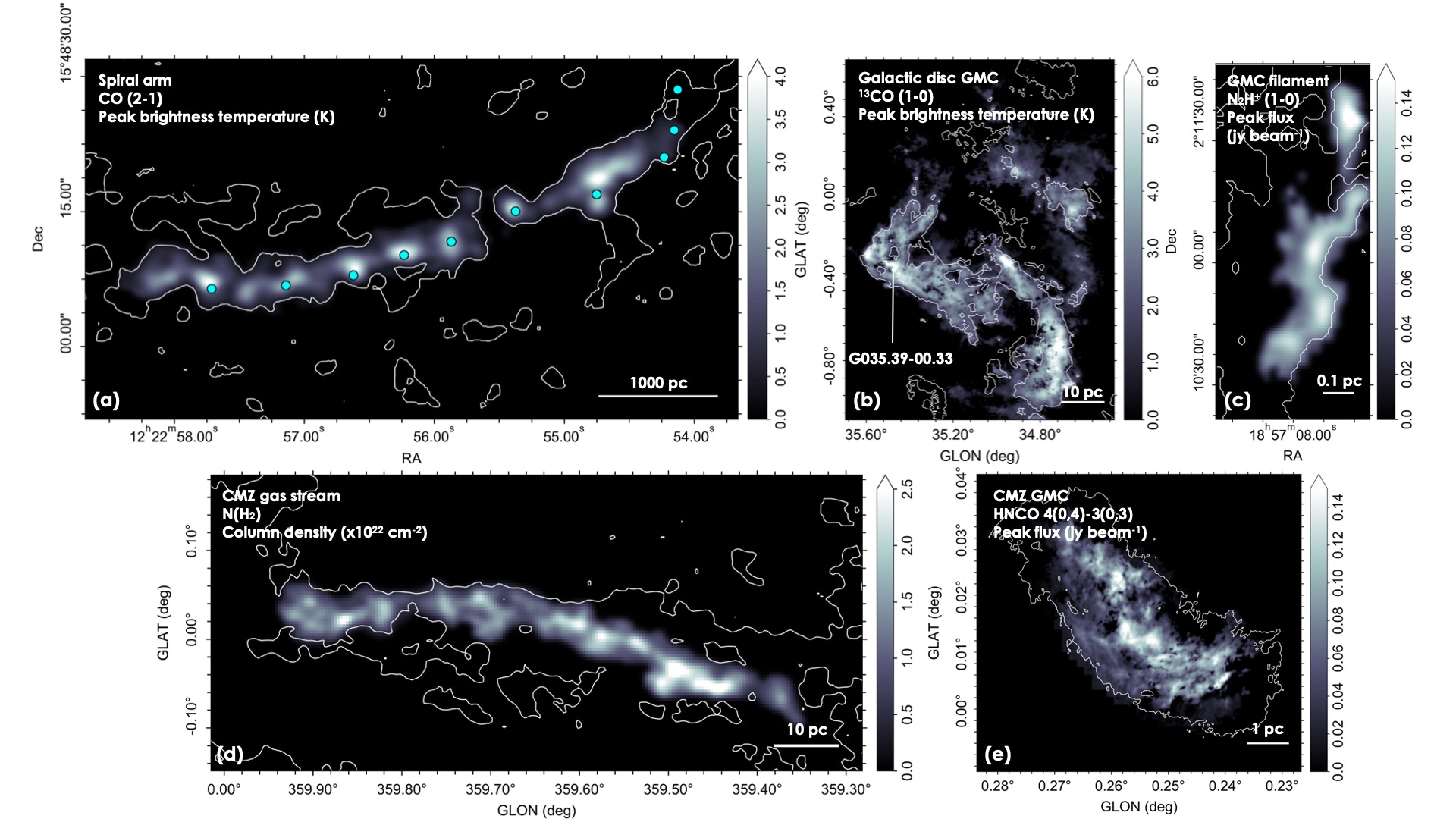}
\end{center}
\vspace{-2mm}
\caption{\label{fig:edfmaps}\textbf{Extended Data Figure 2 $|$ Maps of the regions selected for statistical analysis.} The upper panels display our galactic disc environments. From left to right we show part of the main southern spiral arm in NGC\,4321 (a), a GMC in the Galactic disc (b), and an individual filament located within that same GMC (c). The bottom panels display our selected regions in the CMZ: The series of molecular clouds investigated by Ref.~\citenum{henshaw16c} (d) and an individual GMC located within the CMZ gas stream (e). The cyan points in panel `a' refer to the locations of star forming complexes identified in the mid-infrared\cite{elmegreen18}. In the upper left of each panel we indicate the tracer used to create each image. Scale bars are included in the bottom right corner of each image. These regions correspond to the areas over which we perform our statistical analysis (see {\bf Statistical analysis of the observational data} and \ref{Figure:structfunc}).  }
\vspace{-4mm}
\end{figure*}

\begin{figure*}
\begin{center}
\includegraphics[trim = 0mm 40mm 0mm 40mm, clip, width = 1.0\textwidth]{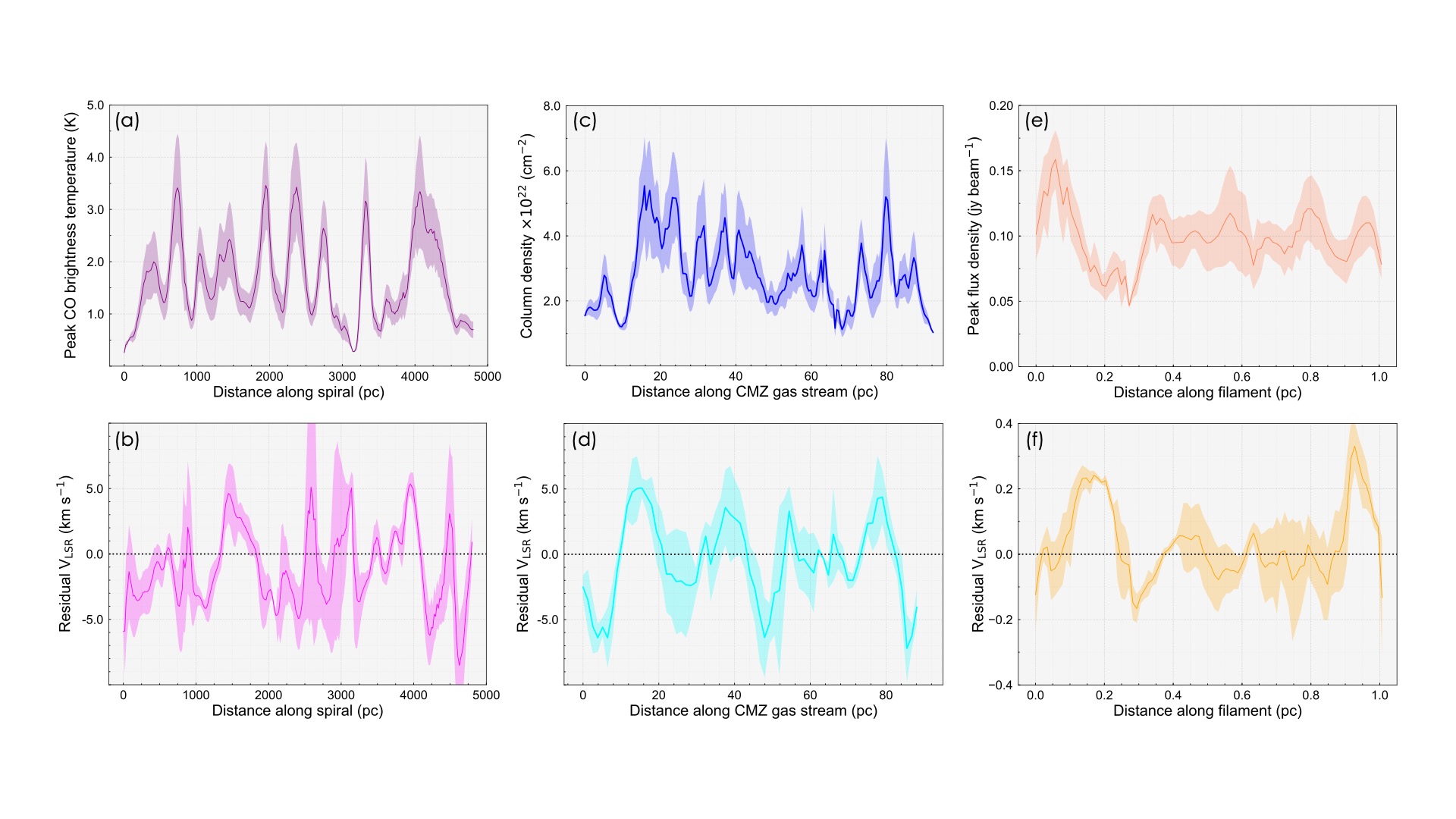}
\end{center}
\caption{\label{fig:edfdenvel}\textbf{Extended Data Figure 3 $|$ Distribution of our density proxy (top) and velocity centroids (bottom) along the crest of the structures displaying periodicity.} From left to right we show distance along the crest of each structure versus mean density (top) and velocity (bottom), for our selected regions in NGC\,4321 (a, b), the CMZ (c, d), and IRDC G035.39--00.33 (e, f), respectively (see \ref{fig:edfmaps}). The coloured shaded region in each panel represents the standard deviation of the data measured orthogonal to the crest.  }
\vspace{-4mm}
\end{figure*}

The emission associated with the GMCs selected in our study do not follow a simple linear morphology (see \ref{fig:edfmaps}b and e). Due to the lack of a dominant geometry, the structure functions presented in \ref{Figure:structfunc}b and e are computed in 2-D, thereby averaging the density and velocity fluctuations over all azimuthal angles in each GMC. 

We generate noise-corrected structure functions by computing the structure function of the measurement uncertainties associated with our Gaussian fit components and subtracting this from the structure functions of the signal (see Supplementary Information). The structure functions presented in \ref{Figure:structfunc}b and e display power-law behaviour over a limited spatial scale, consistent with scale-free fluctuations in both density (dashed lines) and velocity (solid lines). On large-scales, the structure functions begin to flatten due to insufficient sampling\cite{ossenkopf02} (see \ref{Figure:structfunc}). We fit the structure functions with power-law functions of the form $S_{s}(\Bell)\propto\Bell\,^{\zeta_{2}}$. We set the lower and upper limits of our fitting range to the beam size\cite{koch20} (0.65\,pc and 0.07\,pc for the GMCs in the disc and CMZ, respectively) and the approximate scale above which the structure functions begin to turn over (4\,pc and 0.7\,pc for the GMCs in the disc and CMZ, respectively). 

We find that the structure functions of density for our GMCs in the disc and CMZ increase with increasing spatial scale with scaling exponents $\zeta_{2} = {0.82\pm0.03}$ and ${0.91\pm0.01}$. Similarly, we measure scaling exponents for the velocity structure functions of $\zeta_{2} = {0.82\pm0.02}$ and $\zeta_{2}={0.74\pm0.01}$, respectively. 

We also investigate whether the scaling between the velocity structure functions and spatial scale differs as a function of direction. We compute the structure functions as a function of azimuthal angle in 10$^{\circ}$ increments between 0$^{\circ}$ (Galactic east-west) and 90$^{\circ}$ (Galactic north-south). We measure mean scaling exponents of $\zeta_{2} = {0.76\pm0.03}$ and $\zeta_{2}={0.72\pm0.02}$ (where the uncertainty represents the standard deviation of the measurements), for the GMCs in the disc and CMZ, respectively. The small standard deviation of these measurements indicate that the velocity fluctuations have no preferred orientation. We find tentative evidence for a trend between our measured scaling exponents and increasing azimuthal angle in our CMZ GMC ($R^{2}=0.67$, $p$-value$=0.003$). No such trend is evident for our selected GMC in the Galactic disc. 

\clearpage

%%%%%%%%%%%%%%%%%%%%%%%%%%%%%%%%%%%%%%%%%%%%%%%%
%%%%%%%%%%%%%%%%%    FIGS.    %%%%%%%%%%%%%%%%%%
%%%%%%%%%%%%%%%%%%%%%%%%%%%%%%%%%%%%%%%%%%%%%%%%

\begin{figure*}
\begin{center}
\includegraphics[trim = 0mm 120mm 0mm 120mm, clip, width = 1.0\textwidth]{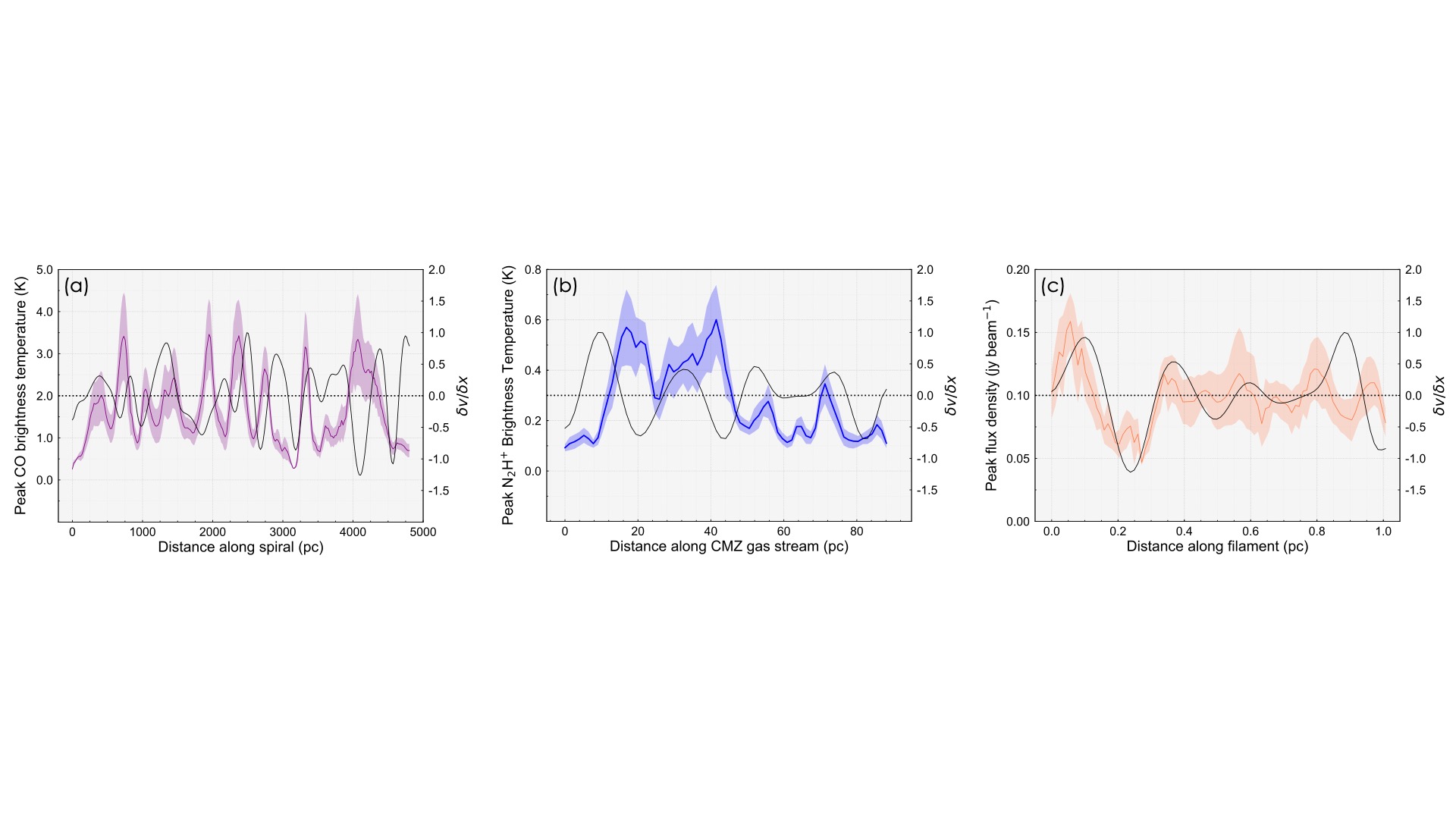}
\end{center}
\vspace{-2mm}
\caption{\label{fig:edfvdiff}\textbf{Extended Data Figure 4 $|$ A comparison between our density proxy and the line-of-sight velocity differential.} Here we show the profile of our density proxy (coloured lines) with the normalised velocity differential (black line) along the crests of our our selected regions in NGC\,4321 (a), the CMZ (b), and IRDC G035.39--00.33 (c), respectively. Note that in panel `b' we show emission of N$_{2}$H$^{+}$ $(1-0)$ rather than the column density distribution displayed in \ref{fig:edfdenvel}c (see the discussion in the Supplementary Information). The black dotted line highlights where the derivative of the velocity is 0.0\,\kms\,pc$^{-1}$. }
\vspace{-4mm}
\end{figure*}

\clearpage

%%%%%%%%%%%%%%%%%%%%%%%%%%%%%%%%%%%%%%%%%%%%%%%%%%%%%%%%%%%%%%%%%%%%%%%%%%%%%%%%%%%%%%%%%%%%%%%%
%%%%%%%%%%%%%%%%%%%%%%%%%%%%%%%%%%      Supplementary Info     %%%%%%%%%%%%%%%%%%%%%%%%%%%%%%%%%
%%%%%%%%%%%%%%%%%%%%%%%%%%%%%%%%%%%%%%%%%%%%%%%%%%%%%%%%%%%%%%%%%%%%%%%%%%%%%%%%%%%%%%%%%%%%%%%%

\clearpage 
\noindent\HorRule\color{black}

\setcounter{figure}{0}
\setcounter{table}{0}
\captionsetup[figure]{labelformat=empty}% redefines the caption setup of the figures environment in the beamer class.
\captionsetup[table]{labelformat=empty}% redefines the caption setup of the figures environment in the beamer class.
\renewcommand{\thefigure}{Supplementary Figure~\arabic{figure}}
\renewcommand{\thetable}{Supplementary Figure~\arabic{table}}

\begin{center}
{\bf \Large \uppercase{Supplementary Information} }
\end{center}

%%%%%%%%%%%%%%%%%%%%%%%%%%%%%%%%%%%%%%%%%%%%%%%%
%%%%%%%%        DECOMPOSITION         %%%%%%%%%%
%%%%%%%%%%%%%%%%%%%%%%%%%%%%%%%%%%%%%%%%%%%%%%%%

\section*{An introduction to spectral decomposition} 

Throughout this work we use the term spectral decomposition to describe the modelling of individual velocity components or emission features detected within spectroscopic data. This is used as an alternative to techniques such as moment analysis, which computes intensity-weighted average quantities, but does not describe the properties of individual emission features where multiple components are evident. Spectral decomposition has been used to study the composition, structure, and dynamics of the ISM, particularly in the Milky Way, where our view through the Galactic plane introduces complex line-of-sight velocity structure\cite{heiles03, hacar13, henshaw14, henshaw16, miville-deschenes17a}.

Spectral decomposition algorithms can be broadly classified into bottom-up and top-down approaches. Bottom-up approaches involve the decomposition of individual spectra independently from one another\cite{ginsburg11, lindner15, clarke18, riener19, sokolov19}. With this approach, additional steps must be taken to ensure (physically-motivated) continuity between adjacent pixels. The top-down methodology involves modelling spatially averaged (and thereby degraded spatial resolution) data, and applying these solutions to the native resolution data\cite{henshaw16, henshaw19, marchal19}. This methodology negates the need for further spatial fitting since it is inherent to the technique.

To decompose the data described in {\bf Observations and data} (Methods), we first use {\sc ScousePy}\cite{henshaw19}, which is a semi-automated spectral decomposition algorithm that falls into the top-down category discussed above. Briefly, {\sc ScousePy} decomposes the data by first breaking a map up into small regions referred to as spectral averaging areas (SAAs). User-provided inputs are used to fit spatially-averaged spectra taken from each SAA. The best-fitting solutions to the SAA spectra are then used as free-parameter guides to a fully-automated fitting procedure which targets the individual spectra located within each SAA. Spatial overlap between SAAs ensures that the fitting process is both non-restrictive and that it is able to transition smoothly between different regions within a data set. This methodology is used for our decomposition of the NGC\,4321, CMZ, G035.39--00.33 and G0.253+0.016 data sets (see \textbf{Spectral decomposition} in Methods). 

The size and complexity of the GRS data set means that a different approach is required. Here, we employ the bottom-up methodology of {\sc GaussPy+}\cite{riener19}, which is specifically designed for the fitting of survey data. {\sc GaussPy+} uses the autonomous Gaussian decomposition algorithm {\sc GaussPy}\cite{lindner15}. {\sc GaussPy} is a machine learning algorithm, originally used in the analysis of H{\sc i} data\cite{lindner15}, which uses a technique known as derivative spectroscopy (see also Ref.~\citenum{clarke18}) to decompose spectroscopic data into individual emission features. {\sc GaussPy+} bolsters the autonomous Gaussian decomposition of {\sc GaussPy} with physically-motivated quality controls and spatially-coherent fitting. For more details on both techniques, we refer the reader to Refs.~\citenum{henshaw16, henshaw19, riener19} and \citenum{lindner15}. 

%%%%%%%%%%%%%%%%%%%%%%%%%%%%%%%%%%%%%%%%%%%%%%%%
%%%%%%%%%%    STRUCTURE FUNCTIONS     %%%%%%%%%%
%%%%%%%%%%%%%%%%%%%%%%%%%%%%%%%%%%%%%%%%%%%%%%%%

\section*{Structure functions and their application to toy models}

Our primary statistic for analysis is the structure function. The structure function of order $p$, relating to a quantity $x$ (for example density or velocity), is defined
\begin{equation}
    S_{p}(\Bell)=\langle\delta x(\mathbf{r},\mathbf{\Bell})^{p}\rangle = \langle |x(\mathbf{r})-x(\mathbf{r}+\Bell)|^{p}\rangle.
\end{equation}
The quantity $\delta x(\mathbf{r},\mathbf{\Bell})$ represents the absolute difference in the quantity $x$ measured between two locations separated by $\Bell$. The structure function, $S_{p}(\Bell)$, averages this quantity (raised to the $p$th power, the order) over all locations (indicated by the angle brackets). The order of the structure function can be directly related to physical quantities. In the case of the velocity structure function, the first order structure function, $S_{1}$, gives an indication of the magnitude of the relative velocity measured throughout a flow on different spatial scales. The second order velocity structure function, $S_{2}$, is proportional to the kinetic energy within the flow. 

The scale-free behaviour of density and velocity fluctuations in turbulent flows, i.e. $\langle\delta x(\mathbf{r},\mathbf{\Bell})^{p}\rangle\propto\Bell\,^{\zeta_{p}}$, means that the structure function is commonly used in the characterisation of interstellar turbulence\cite{kolmogorov41,padoan03,heyer04,chira19}. Structure functions have also been employed in the analysis of time-series data\cite{cordes85, simonetti85}. A useful property of the structure function is that it is sensitive to periodicity in data\cite{lachowicz06, huang10}. The structure function of a periodic quantity will show minima located at the wavelength of the function, as well as at integer multiples of that wavelength, which we demonstrate below. This makes structure functions sensitive to spatial periodicity generated by instabilities\cite{elmegreen83, tafalla15}. In the following sections we devise a series of toy models to help describe the behaviour of the structure function and its defining qualities.

The theoretical formulation of structure functions is based on their application to the density and velocity fields of the fluids, but we are applying the same tools to observational estimators of emission and line centroid velocity.  These are not direct estimators of the underlying fluid fields though observational studies frequently treat them as such.  However, since the velocity centroid is an emission weighted estimate of the velocity field, features in the structure function may be altered by the interaction between the velocity and the density fields. In essence, the velocity field is sampled by the density field and the sampling could potentially alter the inferred structure function. Similarly, instrumental noise may influence the slope of the structure function. Such effects are seen in slopes of power spectrum studies\cite{lazarian00}, but it is not clear whether they will bias the recovered scales in the structure function.  We evaluate some of these effects in the following sections.

%%%%%%%%%%%%%%%%%%%%%%%%%%%%%%%%%%%%%%%%%%%%%%%%
%%%%%%%%%%         TOY MODELS         %%%%%%%%%%
%%%%%%%%%%%%%%%%%%%%%%%%%%%%%%%%%%%%%%%%%%%%%%%%

\subsubsection*{1-D models} 

\begin{figure*}
\begin{center}
\includegraphics[trim = 0mm 30mm 0mm 30mm, clip, width = 0.65\textwidth]{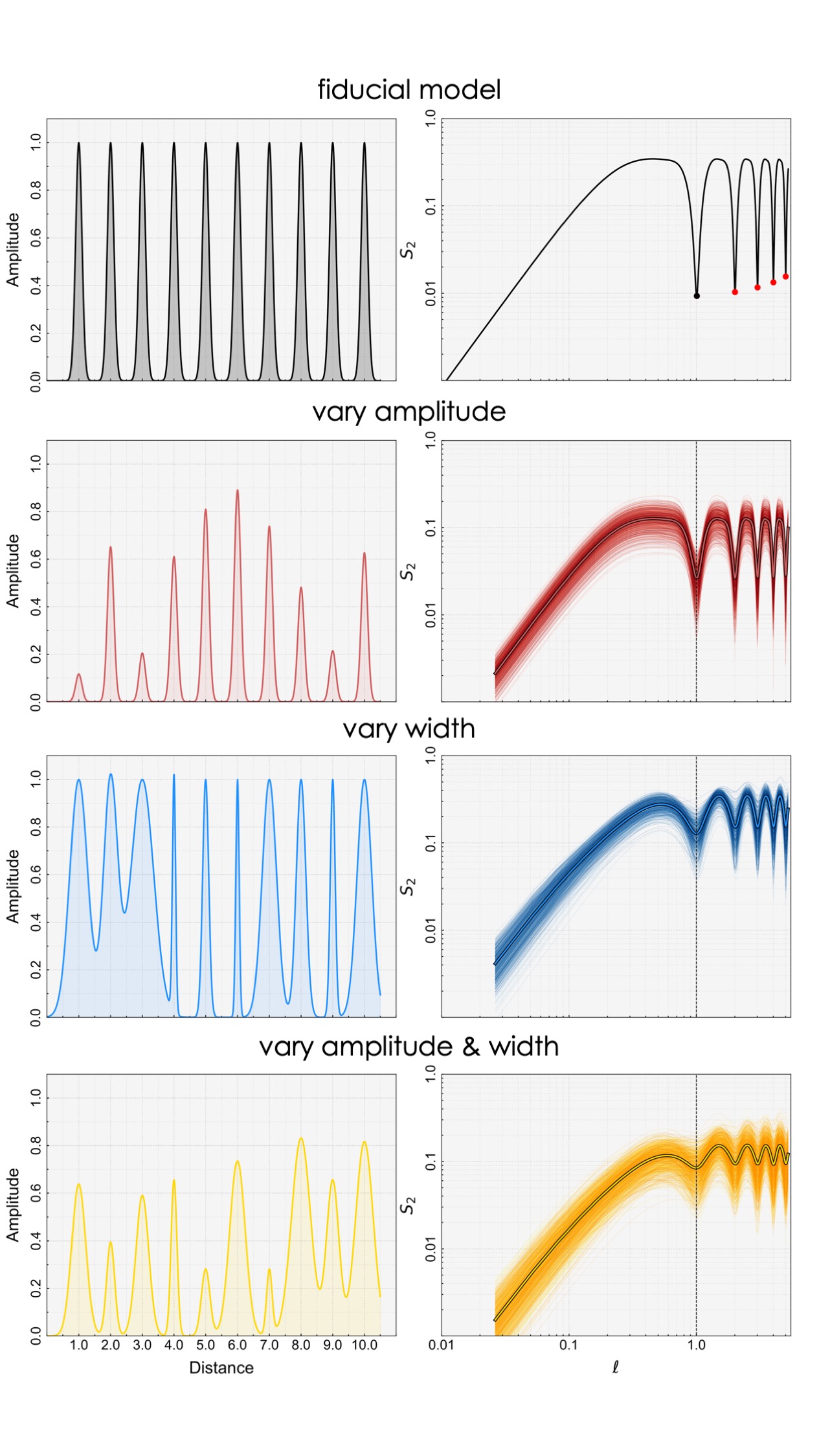}
\end{center}
\vspace{-2mm}
\caption{\label{fig:edf1}\textbf{Supplementary Figure 1 $|$ Gaussian toy models and their structure function response.} The left panels are example cases of toy models analogously representing the density field of regularly-spaced molecular clouds or cores embedded within a spiral arm or filament, respectively. The right hand panels display the structure function of the models on the left. The top panels reflect our fiducial model, 10 equidistant Gaussian peaks with equal amplitude and width. The black circle centred on the first minimum in the top-right panel is located at a lag corresponding to the separation of the Gaussians. The remaining red circles indicate integer multiples of this value. In the remaining panels, from top to bottom we vary the amplitude, width, and both amplitude and width. The narrow individual lines in the right-hand panels correspond to individual structure functions for 1000 realisations of randomly varying the properties of the Gaussians. The thick line corresponds to the average structure function.  The vertical dashed line in each panel represents the median location of the identified minima.   }
\vspace{-4mm}
\end{figure*}

\emph{1-D Gaussian peaks.} Here we investigate the behaviour of the structure function when applied in 1-D to a series of Gaussian peaks. One may think of this toy model as an analogy to the 1-D density structure of a series of regularly-spaced molecular clouds along a spiral arm or pre-stellar cores along a filament. Our fiducial model is displayed in the upper left panel of \ref{fig:edf1} and consists of 10 Gaussians of equal amplitude and width with equidistant spacing. In the upper right panel is the corresponding second order ($p=2$) structure function. The structure function rises to a point of maximum variance after which there is a sharp drop. This drop in $S_{2}$ highlights the separation between Gaussians in the fiducial model, and can be identified in the upper right panel as the first of the minima, occurring at small scales and highlighted by the black circle. The additional minima are located at integer multiples of this spacing (red circles). 

In the remaining panels of \ref{fig:edf1} we highlight the effect of varying the amplitude (upper centre), dispersion (lower centre), and both amplitude and dispersion (bottom) of the Gaussians on the structure function. In each case we keep the spacing between Gaussians fixed at a value of 1.0 (the units are arbitrary), equivalent to that in the fiducial model. In the case of amplitude variation, and where we vary both the amplitude and the width, we allow the amplitude of each Gaussian to vary independently and randomly over the range $0.1-1.0$. Where the width is modified, we allow the dispersion of each Gaussian to vary independently and randomly over the range $0.02-0.4$ (the fiducial value is $0.1$). We then produce 1000 models and estimate the structure function in all cases (individual lines in each right hand panel). The structure function denoted with the thick line represents the average of all experiments in each panel. We identify the location of the first minimum in each of the 1000 cases and find mean values of $1.0$, $1.0$, and $1.25$, respectively. Note that the discrepancy between the latter model and the input spacing is due to the detection of higher order multiples as the ``first minimum'', which is caused by the combination of varying amplitude and width. The median value is $1.0$ in all cases. We conclude that neither amplitude nor dispersion variations have a strong effect on the structure functions' ability to emphasise periodicity, if it exists.

\begin{figure*}
\begin{center}
\includegraphics[trim = 0mm 30mm 0mm 170mm, clip, width = 0.65\textwidth]{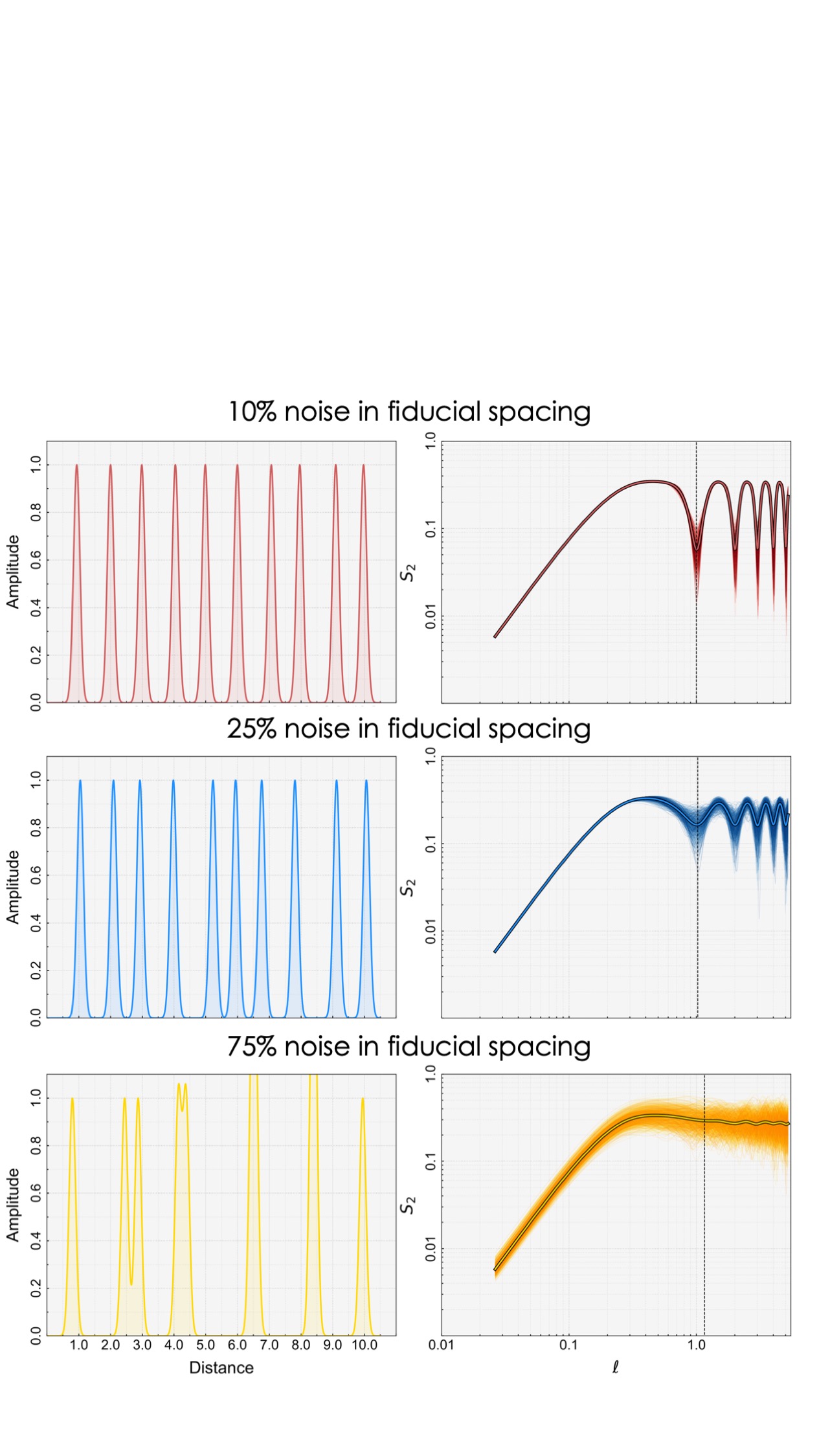}
\end{center}
\vspace{-2mm}
\caption{\label{fig:edf2}\textbf{Supplementary Figure 2 $|$ Investigating the effect of adding noise to the spacing.} In this figure we illustrate the effect of adding noise to the spacing of the individual Gaussian components in the fiducial model displayed in \ref{fig:edf1}. From top to bottom we allow the position of each Gaussian to vary by $\pm$10\%, 25\%, and 75\% of the fiducial spacing. The right hand panel shows the structure function response. The thick and thin lines have equivalent meaning to those presented in \ref{fig:edf1}. The vertical dashed line in each panel represents the median location of the identified minima (see {\bf Structure functions and their application to toy models}). Note that removing integer multiples of the true spacing becomes more difficult as the variation in the locations of the peaks increases and so the location of the minimum begins to diverge from the fiducial spacing. }
\vspace{-4mm}
\end{figure*}

In \ref{fig:edf2} we investigate the effect of adding noise to the location of the Gaussians in the fiducial model. In the examples shown in \ref{fig:edf2} we keep the amplitude and the width of the Gaussians fixed. In the top, centre, and bottom panels we allow the position of each Gaussian to vary with respect to its original location in the fiducial model by $\pm10$\%, $\pm25$\%, and $\pm75$\% $\times$ the fiducial spacing ($=1.0$). We find median locations for the minima (interquartile ranges; IQR) of $1.01 (0.03)$, $1.03 (0.18)$, $1.16 (0.61)$, respectively. Varying the amplitude (between $0.1-1.0$) in addition to the position returns $1.01 (0.03)$, $1.03 (0.26)$, $1.16 (0.61)$. Varying the width (between $0.02-0.4$) in addition to the position returns $1.01 (0.05)$, $1.05 (0.91)$, $1.97 (1.26)$. Finally, varying amplitude, width, and the position returns $1.00 (0.08)$, $1.26 (1.53)$, $2.03 (1.37)$. This analysis indicates that having a variable width in addition to variable position leads to a decreasing likelihood that the fiducial spacing will be identified, which is reflected in the flattening of the structure function displayed in the bottom right panel of \ref{fig:edf2}. The reason for this flattening is that the noise in the location of the Gaussians in the latter model is so large that either no periodicity exists or that the periodicity simply differs from that in the fiducial case (with moderate noise, $\pm25$\%, the input periodicity is still recoverable). This latter point is important. The fact that the fiducial spacing cannot be identified in the bottom right panel of \ref{fig:edf2} is not a reflection on the structure function's ability to determine periodicity more generally. If periodicity exists within data, even if it is generated randomly, it will be seen as an oscillation in the structure function. The bottom right panel of \ref{fig:edf2} should instead serve as a demonstration that care should be taken when interpreting the physical origins of periodicity in real data (cf. Ref.~\citenum{emmanoulopoulos10}). 

\begin{figure*}
\begin{center}
\includegraphics[trim = 0mm 30mm 0mm 30mm, clip, width = 0.65\textwidth]{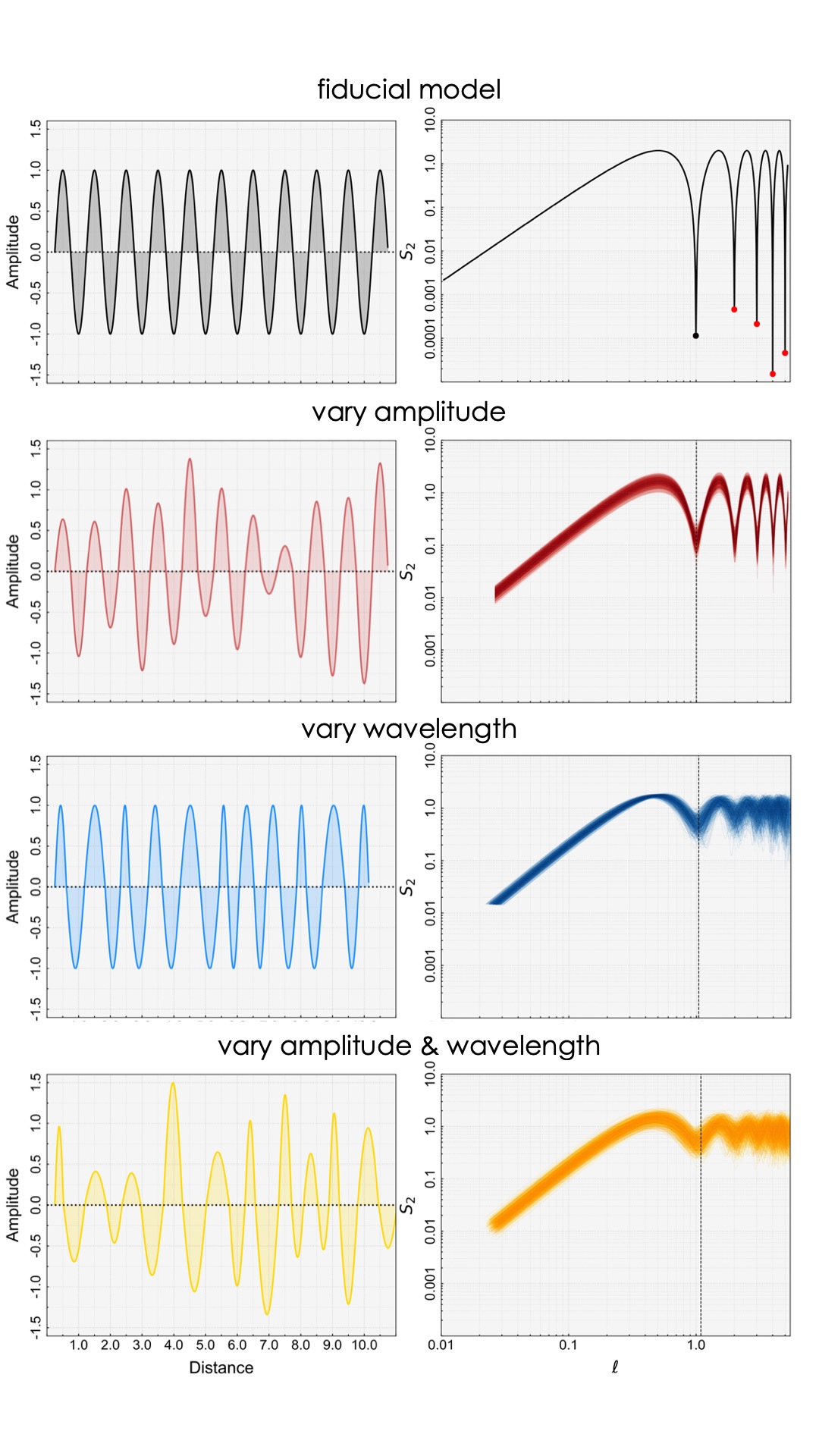}
\end{center}
\vspace{-2mm}
\caption{\label{fig:edf3}\textbf{Supplementary Figure 3 $|$ Sinusoidal toy models and their structure function response.} The left panels are example cases of toy models analogously representing a fluctuating velocity field extracted using spectral decomposition. The right hand panels display the structure function of the models on the left. The top panels reflect our fiducial model, a sine wave with fixed amplitude and wavelength. In the remaining panels, from top to bottom we vary the amplitude, wavelength, and both amplitude and wavelength. The narrow individual lines in the right-hand panels correspond to individual structure functions for 1000 realisations of randomly varying the properties of the signal. The black circle centred on the first minimum in the top-right panel is located at a lag corresponding to the wavelength of the sinusoid. The following red circles indicate integer multiples of this value. The vertical dashed line in each panel represents the median location of the identified minima.}
\vspace{-4mm}
\end{figure*}

\noindent \emph{1-D sine wave.} Here we investigate the behaviour of the structure function when applied in 1-D to a sine wave. We use this in analogy to the centroid velocity field extracted using spectral decomposition. Our fiducial model is displayed in the upper left panel of \ref{fig:edf3}. The fiducial model is a sine wave with amplitude and wavelength of 1.0 (again with arbitrary units). The corresponding structure function is shown in the upper right panel and displays a pronounced minimum at the input wavelength of the fiducial model and further minima located at integer multiples of the wavelength. 

In the lower panels of \ref{fig:edf3} we investigate the effect of irregularity in the amplitude and wavelength of the model on the structure function. To achieve this, we generate $n$ independent sine waves of randomly varying amplitudes and/or wavelengths. We then cut the initial peak of each independent sine wave and stitch these together to form a single coherent toy model with randomly varying properties. In the upper centre, lower centre, and bottom panels we demonstrate the effect of a variable amplitude, wavelength, and both amplitude and wavelength). In the case of amplitude variation, and where we vary both the amplitude and the wavelength, we allow the amplitude to vary independently and randomly over the range $0.25-1.5$ (the fiducial value is $1.0$). Where the wavelength is modified we allow it to vary independently and randomly over the range $0.5-1.5$ (the fiducial value is $1.0$).

The effect on the structure function for each model is displayed in the right hand panels. Here we have performed each experiment a total of 1000 times. The overall effect of the amplitude variation is to reduce the magnitude of the identified dip. For the other two cases, the location of the dip moves (as expected for a variable wavelength). This has the effect of `blurring' or `smearing' the structure function. However, a pronounced dip remains evident with a median location of 1.0. These experiments, although idealised, clearly demonstrate the sensitivity of the structure function to periodicity. 

\subsubsection*{2-D models} 

We use mock data to explore the extent to which scales recovered by structure function analysis are affected by using observational estimators of the density and velocity fields, namely the brightness and velocity centroid.  We create mock density and velocity fields with a power-law power spectrum of fluctuations $P(k)\propto k^\alpha$ and random phase over a periodic volume. From these mock density and velocity fields, we generate mock column density and velocity centroid maps and calculate the structure function for these mock data assuming the emission is optically thin (model `a').  We optionally amplify the power spectrum of either the density or velocity spectrum in a range of wave numbers corresponding to injecting structure on a fixed scale. We are particularly interested in: i) whether a characteristic scale in the density will create an apparent scale in the velocity centroid even if there is no associated scale in the velocity field (model `b'); ii) whether a characteristic scale in the velocity field will create an apparent characteristic scale in the column density (model `c'); iii) whether the structure function applied to observed data can recover (potentially different) scales in the velocity and density fields in their corresponding velocity centroid and column density maps (models `d' and `e'). These models are presented in \ref{fig:edf2d}.

In all these tests, we find that the characteristic scales recovered by the  structure function are not biased by applying it to observational estimators rather than the underlying fields, even though the slopes of the structure functions can be affected by the interplay between the density and velocity fields.  We note that these are simple mock data models without underlying physical basis and no treatment of excitation conditions or optical depth effects.  However, they do not show that using the observational estimators is fundamentally flawed.

\subsubsection*{The impact of noise} 

Instrumental noise introduces measurement uncertainties and spurious fluctuations in the emission extracted from \ppv-data cubes. These fluctuations have the potential to reduce the magnitude of any correlations present in the data\cite{dickman85}.

\begin{figure*}
\begin{center}
\includegraphics[trim = 0mm 200mm 0mm 0mm, clip, width = 0.9\textwidth]{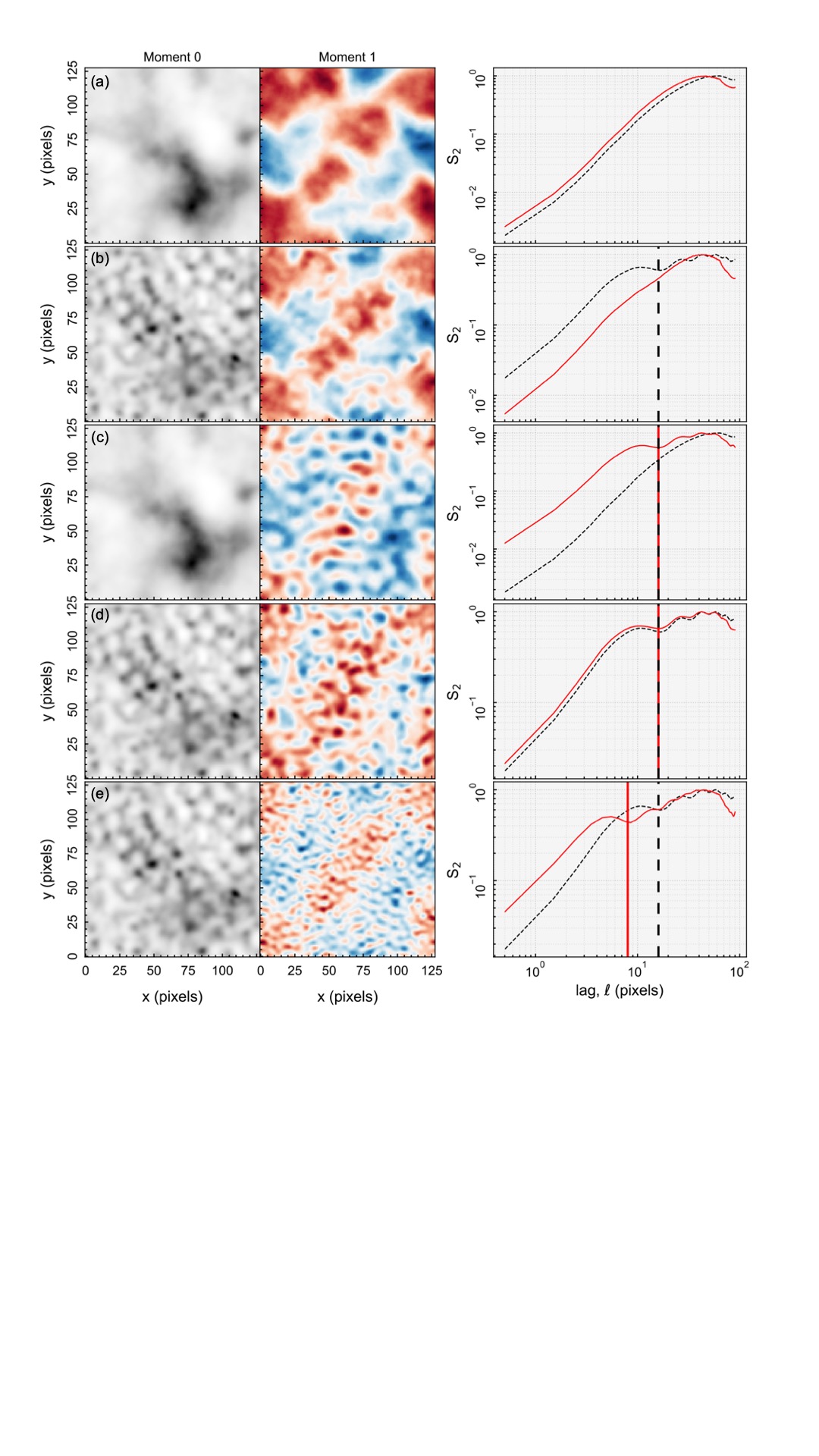}
\end{center}
\vspace{-2mm}
\caption{\label{fig:edf2d}\textbf{Supplementary Figure 4 $|$ Exploring potential sources of bias in using observational estimators.} Mock column density (moment 0; left panels) and velocity centroid (moment 1; centre panels) maps and their corresponding structure functions (in black and red, respectively; right panels). Model `a' shows the fiducial case: column density and centroid velocity maps generated from density and velocity fields with a power-law power spectrum of fluctuations and a random phase over a periodic volume. In model `b', we amplify the power spectrum of density in a range of wave numbers, thereby injecting structure at a fixed scale. In model `c', we do the same but for the velocity power spectrum. In model `d', we inject structure at the same fixed scale in both density and velocity. Finally, in model `e', we inject density and velocity structure on different scales. These tests show that the characteristic scales recovered by the structure function are not biased by applying it to observational estimators of the underlying density and velocity fields.  }
\vspace{-4mm}
\end{figure*}

To evaluate the impact of noise on the structure function and the recovered scales, we introduce Gaussian noise to the models discussed in the previous sections. First, we create a simple 1-D model describing periodic Gaussian emission features. To this model we add a small amount of noise to the spacing of each Gaussian (of the order $\pm5\%$ of the fiducial spacing of 1.0). We allow the amplitude of each Gaussian to vary randomly between $0.3-1.0$, and we allow the width of each Gaussian to vary randomly between $0.05-0.3$. This ensures that there exists some uncertainty in the periodicity of the Gaussian features (cf. \ref{fig:edf1} and \ref{fig:edf2}). Second, we evaluate the effect of noise on the structure function computed for the moment 0 map of model A (in 2-D) described above (see the top left panel in \ref{fig:edf2d}). In both the 1-D and 2-D cases, we add Gaussian noise with a mean of 0.0 but with increasing dispersion. We increase the dispersion from $0.0$ to $\langle I \rangle$ in steps of $0.1\times\langle I \rangle$, where $\langle I \rangle$ represents the mean signal in the model. The presence of noise will change the measured slope of the structure function estimator.   

\begin{figure*}
\begin{center}
\vspace{-2mm}
\includegraphics[trim = 60mm 90mm 60mm 90mm, clip, width = 1.0\textwidth]{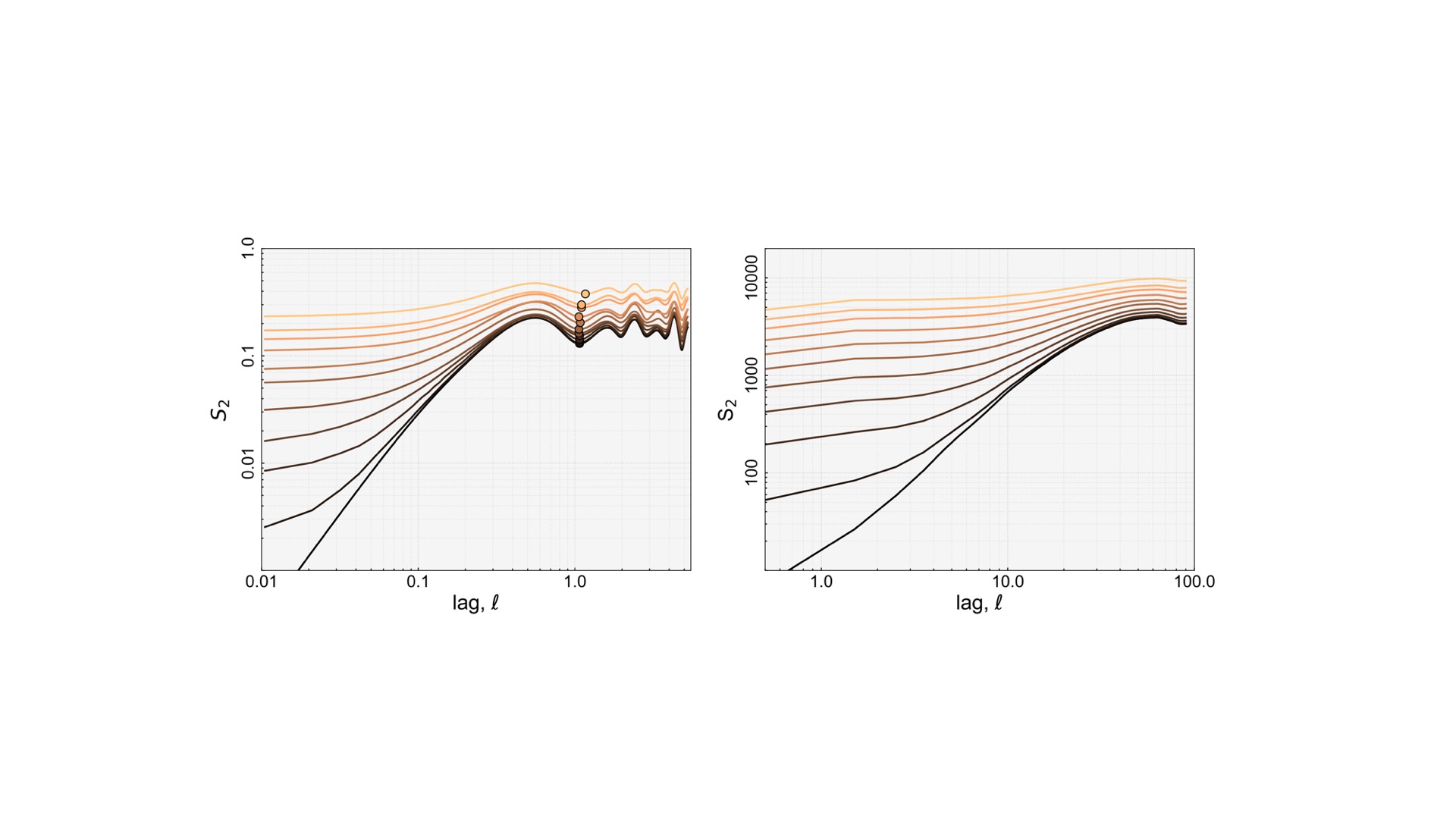}
\end{center}
\caption{\textbf{Supplementary Figure 5 $|$ Testing the impact of noise on the structure function.} Left: Structure functions computed for 1-D models of quasi-periodic Gaussian features, with varying amplitudes and widths. The circles indicate the locations of the minima, corresponding to the periodicity of the Gaussian peaks identified in each case. Right: Structure functions computed for a 2-D synthetic column density map generated from a power-law distribution of density fluctuations. In each case, the structure functions from dark to light show the impact of increasing levels of white noise (the black structure function corresponds to the noise-free case). The maximum noise level in each has a standard deviation equivalent to the mean of the signal in the model.  }
\vspace{-4mm}
\label{fig:edfnoise}
\end{figure*}

The results of this analysis are shown in the top panels of \ref{fig:edfnoise}. In general we see that the structure function tends to flatten as a function of increasing noise level. This makes sense since the structure function of pure white noise is flat. In the case where periodicity is evident, the location of the minimum in each model tends to larger values with increasing noise. The locations of the minima increase from $1.07^{+0.28}_{-0.21}$ to $1.17^{+0.26}_{-0.24}$. The uncertainty here represents the extent of the full-width at half-\emph{minimum} of each local minimum. In general these results demonstrate that while the slope of the structure function is affected by noise, the location of the identified minima are less susceptible to increasing noise levels. 

Given the influence noise has on the profile of the structure function, it is worthwhile exploring whether we are able to correct our observed structure functions for the effects of instrumental noise. To do this, we follow the formalism outlined in Refs.\citenum{dickman85,miesch94,brunt02}. The correction to the structure function can be deduced by first considering its relationship to the autocorrelation function. The autocorrelation function of an arbitrary field is given by
\begin{equation}
    C(\Bell)=\langle x(\mathbf{r})x(\mathbf{r}+\Bell)\rangle,
    \label{eq:acf}
\end{equation}
where the angle brackets denote averaging over all points separated by a lag, $\Bell$. The autocorrelation function is often used in normalised form\cite{dickman85,miesch94}, which we will denote $\overline{C}=C(\ell)/C(0)$ (where $C(0)$ represents the autocorrelation measured at $\Bell=0$), to differentiate it from the form presented in Equation~\ref{eq:acf}. 

The second order ($p=2$) structure function presented in Equation~\ref{eq:sf} can be related to the autocorrelation function by 
\begin{equation}
    S_{2}(\Bell)=\langle(x(\mathbf{r}))^{2}\rangle + \langle(x(\mathbf{r}+\Bell))^{2}\rangle - 2\langle x(\mathbf{r})x(\mathbf{r}+\Bell)\rangle.
\end{equation}
The term on the right is simply the autocorrelation function. For a statistically homogeneous and isotropic field, $\langle (x(\mathbf{r}))^{2} \rangle\approx \langle (x(\mathbf{r}+\Bell))^{2}\rangle \approx \langle (x)^{2} \rangle$, which is simply the autocorrelation function measured at zero-lag, $C(0)$. Hence
\begin{equation}
    S_{2}(\Bell)=2[\langle (x)^{2} \rangle - C(\Bell)]=2[C(0)-C(\Bell)]
\end{equation}

The effect of noise on the structure function is demonstrated in the top panels of \ref{fig:edfnoise}. The effect of noise is most prominent at small lags, leading to the flattening of the structure function. To explore how this behaviour may be compensated for, we denote by $x(\mathbf{r}$) and $x^{\prime}(\mathbf{r}$) the arbitrary field measured in the presence and absence of noise, respectively. If the uncertainty in the measurement is denoted $\delta x(\mathbf{r})$ then we can write
\begin{equation}
    x(\mathbf{r})=x^{\prime}(\mathbf{r}) + \delta x(\mathbf{r}).
\end{equation}

Expressing the autocorrelation function presented in Equation~\ref{eq:acf} in terms of the noise-free measurements
\begin{equation}
    C(\Bell)=\langle [x^{\prime}(\mathbf{r})+\delta x(\mathbf{r})][x^{\prime}(\mathbf{r}+\Bell)+\delta x(\mathbf{r}+\Bell)]\rangle.
\end{equation}
and by expanding we find
\begin{equation}
    C(\Bell)=C^{\prime}(\Bell) + C_{n}(\Bell) + \langle x^{\prime}(\mathbf{r})\delta x(\mathbf{r}+\Bell)+ x^{\prime}(\mathbf{r}+\Bell)\delta x(\mathbf{r}) \rangle.
\end{equation}
Here, the first two terms represent the autocorrelation functions of the noise-free field and the noise itself, respectively. The terms in the angle brackets represent the correlations between signal and noise. For random, Gaussian noise we can assume that the signal and noise are uncorrelated and treat this term as zero\cite{brunt02}, thus
\begin{equation}
    C(\Bell)=C^{\prime}(\Bell)+C_{n}(\Bell).
\end{equation}

Since Gaussian noise is uncorrelated and has zero mean ($\langle \delta x(\mathbf{r})\rangle=0$), it follows that
\begin{equation}
    C_{n}(\Bell) = \begin{cases}
    \sigma_{n}^{2} &\text{for $\Bell=0$}\\
    0 &\text{for $\Bell\ne0$}
\end{cases}
\end{equation}
where $\sigma_{n}^{2}$ represents the variance of the noise. Combining this information we can see that for non-zero lags, the autocorrelation function remains unchanged by Gaussian noise
\begin{equation}
     C(\Bell)=C^{\prime}(\Bell), \quad \text{for $\Bell\ne0$}.
\end{equation}
However, at zero lag 
\begin{equation}
     C(0)=C^{\prime}(0)+\sigma_{n}^{2}, \quad \text{for $\Bell=0$}.
\end{equation}

The equation for the noise-free structure function of a statistically homogeneous and isotropic field is
\begin{equation}
    S^{\prime}_{2}(\Bell)=2[C^{\prime}(0)-C^{\prime}(\Bell)].
\end{equation}
With the knowledge that $C(\Bell)=C^{\prime}(\Bell)$ and $C^{\prime}(0)=C(0)-\sigma_{n}^{2}$ at non-zero and zero lags, respectively, we can write
\begin{equation}
    S^{\prime}_{2}(\Bell)=2[C(0)-C(\Bell)-\sigma_{n}^{2}],
\end{equation}
and hence
\begin{equation}
    S^{\prime}_{2}(\Bell)=S_{2}(\Bell)-2\sigma_{n}^{2},
    \label{eq:sfnoisecorrsig}
\end{equation}
which is equivalent to the result presented in Ref.~\citenum{miesch94} (although Ref.~\citenum{miesch94} present the result for the normalised structure function). 

\begin{figure*}
\begin{center}
\vspace{-2mm}
\includegraphics[trim = 60mm 0mm 60mm 0mm, clip, width = 1.0\textwidth]{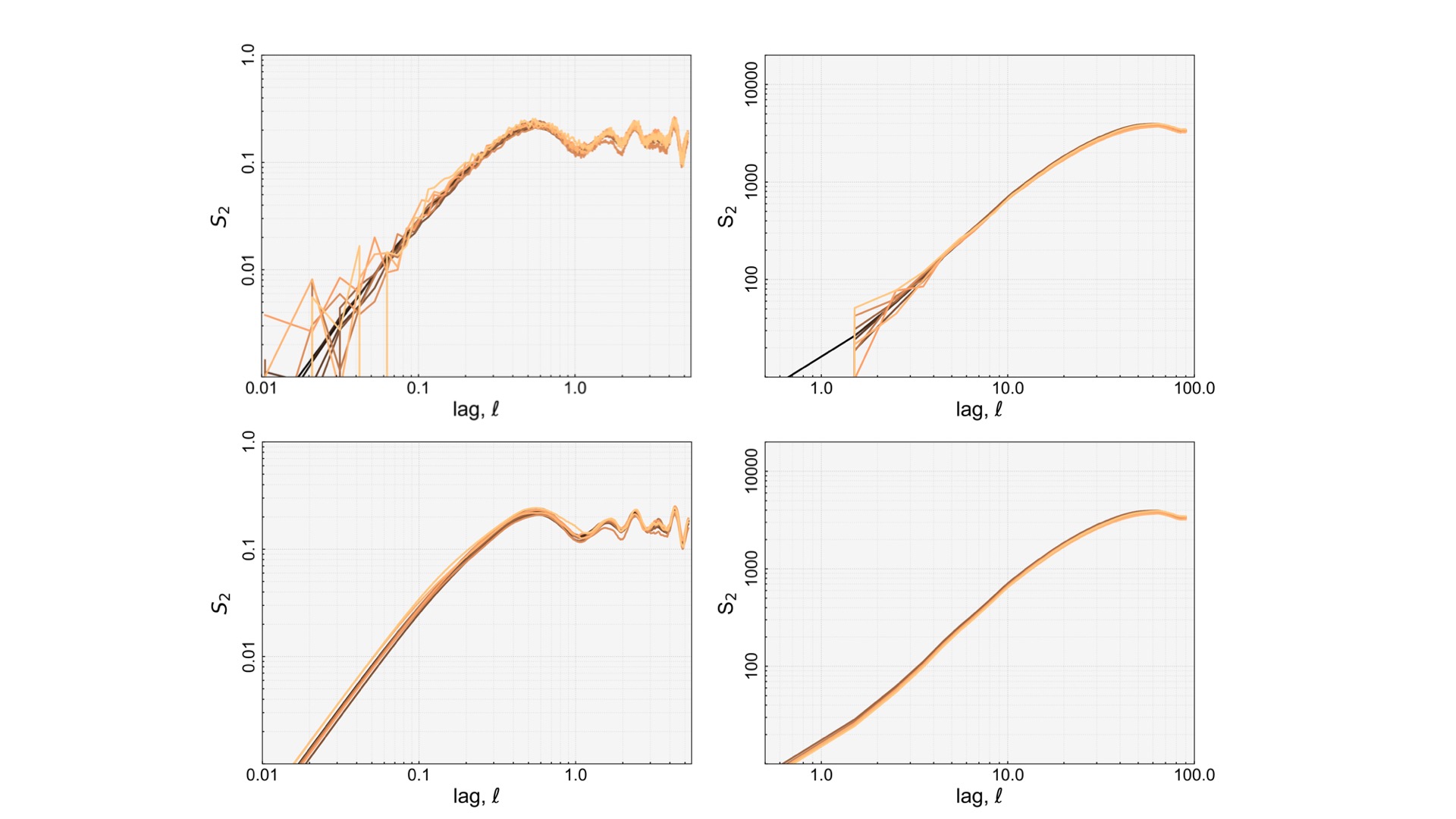}
\end{center}
\caption{\textbf{Supplementary Figure 6 $|$ Noise-corrected structure functions.} Top panels: Noise-corrected structure functions according to Equation~\ref{eq:sfnoisecorrsig}. Bottom panels: Noise-corrected structure functions after subtracting the structure function of the noise as in Equation~\ref{eq:sfnoisecorr}. The left and right panels correspond to the 1-D and 2-D models, respectively. Colours have equivalent meaning to those in \ref{fig:edfnoise}. }
\vspace{-4mm}
\label{fig:edfnoisecorr}
\end{figure*}

In the top panels of \ref{fig:edfnoisecorr} we show noise-corrected structure functions according to Equation~\ref{eq:sfnoisecorrsig}. At large lags we find that this simple correction does an adequate job of removing the noise from the structure function. At small lags the correction is imperfect where the magnitude of the structure function is comparable to the variance of the noise. The impact of the noise is therefore evident at greater lags for higher noise levels. In principle, one can also derive
\begin{equation}
    S^{\prime}_{2}(\Bell)=S_{2}(\Bell)-S_{2,n}(\Bell),
    \label{eq:sfnoisecorr}
\end{equation}
which represents subtracting the structure function of the noise. Observational data will usually have spatial correlations from the beam and scanning effects that will more accurately be accounted for by subtracting the structure function of the noise (which can be computed using Equation~\ref{eq:sf}). We display the results using this method in the bottom panels of \ref{fig:edfnoisecorr}. This modelling demonstrates that the profile of the structure function can be returned with an accurate measurement of the underlying noise distribution.  

%%%%%%%%%%%%%%%%%%%%%%%%%%%%%%%%%%%%%%%%%%%%%%%%
%%%%%%%%% PHYSICAL INTERPRETATION %%%%%%%%%%%%%%
%%%%%%%%%%%%%%%%%%%%%%%%%%%%%%%%%%%%%%%%%%%%%%%%
\section*{Summary of the results and physical interpretation} 

Our key observational findings are as follows. First, we identify ubiquitous velocity fluctuations throughout the molecular interstellar medium. Second, while the velocity fluctuations observed throughout our selected GMCs are scale-free and without a singular orientation, this is not the case in all regions. Along the crests of the filamentary sub-regions selected in this study, the observed velocity fluctuations oscillate with characteristic wavelengths. Third, in the regions where regular velocity oscillations are detected, correlated periodicity is evident in the density structure of the ISM. In these regions, the separation between regularly-spaced density enhancements agrees (within the uncertainties) with the characteristic wavelength of the oscillations detected in velocity. Here we describe the key information which provides the bases for our physical interpretation of the results.

\subsubsection*{Origins of the periodic density fluctuations} 

The separation between density enhancements is well-resolved in each environment. In the spiral arm, the CMZ gas stream, and the GMC filament, the physical separation between density enhancements corresponds to around 3, 5 (18 for the second minimum), and 5 resolution elements, respectively (see {\bf Observations and data} in Methods). Further, the derived separation between the molecular cloud complexes traced by CO ($2-1$) emission in the spiral arm of NGC\,4321 ($366^{+88}_{-77}$\,pc) agrees with the spacing of 410\,pc, derived independently, using tracers of embedded star formation in Ref.~\citenum{elmegreen18}. Our value is also close to the separation extracted in Ref.~\citenum{chevance19} ($248^{+33}_{-26}$\,pc), where the PHANGS-ALMA CO ($2-1$) data is also used. The main difference between our derived separation and that presented in this latter study is that Ref.~\citenum{chevance19} do not focus exclusively on spiral structure. Instead, their structure identification is performed over the entire field of view of the PHANGS-ALMA data. Similarly, the minimum located at $21.8^{+5.5}_{-6.3}$\,pc in the density structure function of the CMZ gas stream, measured using \emph{Herschel} observations (\ref{Figure:structfunc}d), agrees with the minimum in the independently-derived velocity structure function ($22.0^{+5.4}_{-6.3}$\,pc), measured using N$_{2}$H$^{+}$ ($1-0$) observations from Mopra (discussed below in {\bf Origins of the periodic velocity fluctuations}). These independent measurements lead us to conclude that the density enhancements within our selected sub-regions exhibit well-defined separation scales. 

The regular spacing of density enhancements detected in our observations is akin to structure observed in the discs of spiral galaxies, in which periodically-spaced star-forming regions have been analogously compared to `beads on a string'\cite{baade58,elmegreen83}. This periodicity has been noted both in direct tracers of star formation\cite{elmegreen83,la-vigne06,gusev13,elmegreen18,elmegreen19}, as well as in molecular gas\cite{espada12}. Analogously, on small scales, both dust continuum\cite{wang11,kainulainen13b,henshaw16b,liu18,zhang20} and molecular line observations\cite{jackson10,hacar11,tafalla15} of molecular clouds in the Milky Way have revealed quasi-periodically spaced chains of dense clumps and pre-stellar cores along the crests of dense filaments. 

The quasi-periodicity of such density enhancements are often discussed in the context of gravitational instabilities, with numerical work showing that sinusoidal perturbations in density along an infinitely long isothermal cylinder with wavelengths exceeding some critical value will produce an instability in the medium\cite{stodolkiewicz63}. The critical perturbation length scale, in the absence of a magnetic field, is $\lambda_{c}=3.94(2 c_{s}^{2}/\pi G\rho)^{0.5}$, where $c_{s}$ is the thermal sound speed and $\rho$ the gas density. These perturbations will fragment a cylinder into regularly-spaced density enhancements where the separation is given by the wavelength of the fastest growing unstable mode, which is approximately twice that of the critical wavelength\cite{nagasawa87} $\lambda\approx22H=22c_{s}/(4\pi G\rho)^{0.5}$ (where $H$ is the isothermal scale height of the filament), or roughly 4 times the filament diameter\cite{inutsuka92}. The corresponding timescale for the growth of the instability is $t=~2.95(4\pi G\rho)^{-0.5}$. 

On the smallest scales in our study, we show that the periodicity of pre-stellar cores in our GMC filament is of the order $3$ times the deconvolved filament diameter (see {\bf Analysis of filamentary structures} in Methods), consistent with the above framework. We can also turn this question around and ask: what would be the expected central density of a filament required to produce our observed periodic spacing of $0.32$\,pc? If we assume a sound speed of $0.27$\,\kms \ (relevant for cold molecular gas at a temperature of 20\,K), we derive a central density and growth timescale of $10^{5}$\,cm$^{-3}$ and 0.15\,Myr, respectively (note that using our mean measured velocity dispersion of $\sim0.3$\,\kms \ changes the central density by $<30\%$). This density is consistent with the estimated number density of star-forming cores within the cloud\cite{henshaw16b} and with the density of other high-mass filaments observed with similar spatial resolution\cite{peretto13,hacar17}. The consistency between the observed periodicity relative to the filament diameter, as well as between the model predicted central density of the filament and that expect for dense filaments in IRDCs, indicates that the observed structure may be the direct result of gravitational instabilities. 

What about the larger-scale periodic features detected in our observations? Applying the aforementioned framework to the spiral arm of NGC\,4321 and the CMZ gas stream is less trivial. The simplistic model outlined above, which describes the gravitational fragmentation of an isolated cylindrical filament in hydrostatic equilibrium, does not capture the important effects of shear and magnetic fields. Despite this, a key finding, detected in observations\cite{elmegreen83, elmegreen18, elmegreen19} and further evident in numerical simulations of galaxy discs\cite{elmegreen94, kim02, renaud13}, is that the separation of star-forming complexes situated along spiral arms is of the order a few times the diameter of their parent filament. This is qualitatively similar to the periodicity sometimes observed in local filaments (see above), albeit on much larger scales, and is consistent with the separation predicted by the simple model described above. 

Why is it therefore, that the simplistic model of a fragmenting cylinder may serve as a reasonable approximation for describing the observed `beads on a string' morphology of star-forming complexes in nearby galaxy discs? One explanation is that the low levels of shear in the dense inner regions of spiral arms (also in galactic nuclei), and the subsequent reduction of stabilising Coriolis forces, is conducive to the development of instabilities\cite{elmegreen87, elmegreen94, kim02}. Another explanation is that the presence of a magnetic field may actively promote the destabilisation of the gas by opposing the Coriolis force, transferring angular momentum away from growing condensations\cite{elmegreen94, kim02, inoue19}. Whatever the source of the reduced shear, so long as it is efficiently reduced (cf. Ref.~\citenum{inoue18}), the development of gravitational instabilities in the azimuthal direction becomes comparable, modulo a modification of the growth time, to the non-magnetic, non-rotating case\cite{elmegreen87,kim01}. Once initiated, these instabilities are capable of driving the formation of periodic density enhancements\cite{kim02,dobbs08,renaud13}, the characteristic spacing of which is broadly consistent with that predicted by the simplistic model described above and, crucially, with the observations\cite{elmegreen83, elmegreen18}.

The periodic structure evident in our observations of the spiral arm of NGC\,4321 and the CMZ gas stream is consistent with this picture. The derived periodicity in these environments corresponds to $3$ and $5$ times the beam-deconvolved diameter of the respective parent structure (see {\bf Analysis of filamentary structures} in Methods). We conclude from this discussion that gravitational instabilities represent a plausible source of the periodic structure identified in this study, and furthermore speculate that the development of gravitational instabilities across multiple scales in the ISM may be a critical (though not exclusive) ingredient for the formation of dense star-forming gas. 

\subsubsection*{Origins of the periodic velocity fluctuations} 

Our toy models demonstrate that the characteristic scales recovered by the structure function are not biased by applying it to observational estimators rather than the underlying density and velocity fields (see {\bf 2-D models} above). Although simplistic, these models show that periodic fluctuations evident in the density field do not necessarily create an apparent and correlated characteristic scale in the observed velocity centroid, even if there is no associated scale present in the true velocity field. These factors give us confidence that the correlated periodicity between density and velocity is not simply an artefact of the fact that our spectral decomposition only probes observational estimators of the true underlying density and velocity fields. Similar features observed in both low-mass\cite{hacar11, arzoumanian18} and high-mass\cite{liu19} filaments, using independent tracers of gas column-density and centroid velocity (see also the CMZ data presented in this work), instead points to a dynamical origin of the velocity oscillations, strengthening our argument that such motions may be a common feature of the molecular ISM on all scales (\ref{Figure:wiggles}).

What is the underlying physical mechanism driving the observed periodic velocity oscillations? In the previous section we demonstrated the plausibility that gravitational instabilities may create periodic density fluctuations along the crests of our filamentary structures. Our findings now extend this result by showing that the separation-to-diameter ratio in the velocity structure of the ISM matches that of the density enhancements. To assess whether the velocity oscillations are the result of converging flows towards the periodic density enhancements, as would be the case for gravitational contraction, in \ref{fig:edfvdiff} we compare our density proxy (coloured lines) in each region with the normalised velocity differential (black line). If the inferred gas motions are the result of gravitational collapse, we would expect that the locations of density enhancements coincide with extrema in the velocity gradient\cite{clarke16,misugi19}. 

The physical interpretation of the observed gas flows can be further inferred from the phase shift between the density and velocity fluctuations (see {\bf Analysis of filamentary structures} in Methods). If there is no phase difference between the density and velocity fluctuations, this indicates that density enhancements are located at extrema in the measured velocity, and therefore close to where the velocity gradient is zero. Similarly, a phase shift of $\lambda/2$ indicates that density enhancements are situated at either positive or negative extrema in velocity (but not both), and are again located where the velocity gradient is close to zero. Convergent motion is therefore characterised by a small, but non-zero, phase shift between density and velocity, indicating that the density enhancements are located close to extrema in the velocity gradient.

\ref{fig:edfvdiff} shows that there exists a spatial correlation between the centres of mass of the periodic density enhancements and non-zero values of the velocity differential in both the CMZ gas stream (panel `b') and the GMC filament (panel `c'). 

It is worth noting that in \ref{fig:edfvdiff}b we show the density field in the CMZ as traced by N$_{2}$H$^{+}$ ($1-0$) emission. As pointed out in Ref.~\citenum{henshaw16c}, many of the column-density peaks identified in the \emph{Herschel} data of the CMZ are situated close to velocity extrema (both blue and red-shifted), where the velocity differential is close to zero. This indicates that density peaks are more or less half-spaced along the velocity oscillation, which is reflected in the small phase difference between density and velocity of $2.7\pm2.9$\,pc measured via the cross-correlation (see {\bf Analysis of filamentary structures} in Methods). However, an important observation, revealed by our subsequent analysis, is that the structure function of the column-density shows a minimum located at $21.8^{+5.5}_{-6.3}$\,pc, very close to the location of the minimum in the independently-measured velocity structure function ($22.0^{+5.4}_{-6.3}$\,pc). This was not captured in the analysis of Ref.~\citenum{henshaw16c}, who focussed exclusively on the nearest-neighbour separations of the column density peaks (which gives rise to the first minimum located at $6.0^{+0.8}_{-0.6}$\,pc). 

If we instead focus on the N$_{2}$H$^{+}$ $(1-0)$ emission, the peaks located at around 20\,pc, 40\,pc, and 75\,pc in \ref{fig:edfvdiff}b are coincident with extrema in the velocity gradient measured along the crest of the stream. Revisiting the column density profile along the gas stream presented in \ref{fig:edfdenvel}c, we see that the column density peaks are gathered in groups which correlate with the locations of the N$_{2}$H$^{+}$ $(1-0)$ peaks. We therefore propose that the gas stream has first fragmented into clumps with a separation of around 20\,pc (roughly 5 times the diameter of the CMZ gas stream; see \ref{tab:lengthscales} and {\bf Origins of the periodic density fluctuations}), and it is these density enhancements, detected as peaks in N$_{2}$H$^{+}$ $(1-0)$ emission, which are associated with the observed velocity fluctuations. The impact of this fragmentation remains evident in the higher angular resolution \emph{Herschel} column density data, which shows groupings of molecular clouds (\ref{fig:edfdenvel}c), the signature of which appears as a minimum in the structure function located at $21.8^{+5.5}_{-6.3}$\,pc (\ref{Figure:structfunc}d). These clumps may have then fragmented on smaller scales, akin to the nested modes of fragmentation recently detected on smaller scales in e.g. Orion\cite{kainulainen17}, which is evidenced by the minimum in the structure function located at $6.0^{+0.8}_{-0.6}$\,pc (\ref{Figure:structfunc}d). 

The relationship between density and the velocity differential in the spiral arm of NGC\,4321 is notably different (\ref{fig:edfvdiff}a). Although the CO peaks located at around 350\,pc and 4125\,pc have a velocity profile consistent with what we would intuitively expect for longitudinal gravitational collapse (i.e. density peaks coincident with an extremum in the velocity gradient), this is not typical of this system. Rather, many of the CO peaks evident in \ref{fig:edfvdiff}a are located where the velocity gradient is close to zero. Furthermore, many of the CO peaks align with locations where the velocity gradient is rising from negative to positive values as a function of increasing distance along the arm (see peaks at around 750\,pc, 1000\,pc, 2000\,pc, 2375\,pc, 2750\,pc, and 3375\,pc), indicating that the peaks are almost all blue-shifted with respect to the galactic average at this location. This observation is consistent with the $\lambda_{v}/2$ phase difference measured between density and velocity ($191\pm62$\,pc; see {\bf Analysis of filamentary structures} in Methods). 

We interpret the systematic blue-shift of the CO peaks as the result of bulk motion in the gas arising from a combination of spiral streaming motions and gravitational torque-driven radial inflow. The region selected for our study is located within the corotation radius of the spiral pattern (7.1-9.1 kpc\cite{elmegreen89, garcia-burillo98}) and situated on the receding side of the galaxy, which is rotating in the anti-clockwise direction. Given the alignment of the southern arm portion with respect to the line of sight, the radially inward component of gas motion approaching the underlying spiral potential would lead to a systematic blue-shift of velocities at all locations along this portion of the arm. Radial inflow motions driven by negative gravitational torques expected inside corotation would also contribute to the blue-shift. The magnitude of the velocity oscillations (of the order 5\,\kms \ but up to 10\,\kms; \ref{fig:edfdenvel}b) is consistent with the magnitude of the velocity fluctuations observed more broadly on $<500$\,pc scales throughout the PHANGS-ALMA survey of nearby galaxies (detected as superposed fluctuations on galactic rotation curves\cite{lang19}).

We conclude from this discussion that the periodic velocity oscillations are likely the result of convergent motion, either longitudinal convergence as in the case of the CMZ and the GMC filament, or radial convergence as in the case of the spiral arm of NGC\,4321. 

\subsubsection*{Origins of scale-free density and velocity fluctuations} 

In contrast to the periodicity described above, the density and velocity fluctuations observed throughout both of our selected GMCs do not show any characteristic scale. This is evident from the power-law scaling of the structure functions (see \ref{Figure:structfunc}b, e, and {\bf Analysis of multi-dimensional structures} in Methods).

We can relate our structure function of velocity, $\langle\delta v^{p}\rangle\propto\Bell\,^{\zeta_{p}}$, to the more commonly measured velocity dispersion-size relationship, $\delta v\propto\Bell\,^{\gamma}$, where $\gamma=\zeta/p$ and $p=2$, for the second-order structure function. For the GMCs in the Galactic disc and the CMZ we find $\gamma=0.41\pm0.01$ and $0.37\pm0.01$, respectively. These are slightly lower than the mean Galactic velocity dispersion-size relationship scaling exponent of about 0.5\cite{heyer15}, although they comfortably fall within the distribution of measured values\cite{heyer15}. 

The observed scale-free behaviour of our measured density and velocity fluctuations is reminiscent of the gas structure generated by interstellar turbulence\cite{elmegreen04, maclow04}. Indeed, a velocity dispersion-size scaling exponent of around 0.5 is often cited as evidence for compressible turbulent flows in molecular clouds\cite{heyer15}. However, the observed scaling between velocity fluctuations and spatial scale has a number of (mutually exclusive) interpretations. It has been cited as evidence for the global virialisation of molecular clouds\cite{heyer09} as well as the global collapse of clouds\cite{ballesteros-paredes11}, in addition to the universality of supersonic turbulence\cite{larson81}. 

A further source of uncertainty is that our spectral decomposition is sensitive only to observational estimators of the true underlying density and velocity fields. Although we are confident that the readily distinguishable gas in the CMZ\cite{henshaw16, henshaw19} and the separation of our Galactic disc GMC from the tangent point velocity\cite{riener20} help to mitigate the impact of velocity crowding in these sources (see \textbf{Data selection for statistical analysis} in Methods), the fact that velocity centroids extracted from $p$-$p$-$v$ data cubes do not exclusively trace velocity, but a complex convolution of density, velocity, and excitation, remains\cite{lazarian00}. This unavoidable source of uncertainty means that the exact quantitative details of the structure function scaling derived in this study should be interpreted with a degree of caution. Nevertheless, we are confident that the velocity fluctuations detected throughout our selected GMCs exhibit no discernible characteristic scale (see {\bf 2-D models}).

More detailed future examination of the velocity centroids\cite{lazarian00, esquivel05,ossenkopf06} may help to determine the exact origin of the observed velocity fluctuations. Expanding on our study of the anisotropy in the scaling of the velocity structure functions may also help in this regard. Although we find tentative evidence for directional anisotropy in our selected GMC in the CMZ, no such trend is evident in our Galactic disc GMC (see {\bf Analysis of multi-dimensional structures} in Methods). Further investigation is therefore required to determine if the anisotropy in the scaling of our velocity structure functions in our selected GMC in the CMZ can be related to, for example, the orientation of shock fronts\cite{chira19} or magnetic field lines\cite{heyer08,hu19}. 

%%%%%%%%%%%%%%%%%%%%%%%%%%%%%%%%%%%%%%%%%%%%%%%%
%%%%%%%%%%%%%%%%%   MOVIES.   %%%%%%%%%%%%%%%%%%
%%%%%%%%%%%%%%%%%%%%%%%%%%%%%%%%%%%%%%%%%%%%%%%%

\section*{Supplementary movies} 

We provide Supplementary Videos of the $p$-$p$-$v$ decomposition of the data shown in \ref{Figure:wiggles}. We also provide $p$-$v$ representations of the same data. These videos clearly highlight the velocity fluctuations evident throughout each environment. Note that the $p$-$v$ representation of the Galactic disc includes the entire decomposition of the GRS data set\cite{riener20}. This video demonstrates that the fluctuations are evident across the entire region of the Galactic disc probed by the GRS, and not just the region selected for our study.

%%%%%%%%%%%%%%%%%%%%%%%%%%%%%%%%%%%%%%%%%%%%%%%%
%%%%%%%%%%%%%%%%%    FIGS.    %%%%%%%%%%%%%%%%%%
%%%%%%%%%%%%%%%%%%%%%%%%%%%%%%%%%%%%%%%%%%%%%%%%

%%TC:endignore

\newpage

%%%%%%%%%%%%%%%%%%%%%%%%%%%%%%%%%%%%%%%%%%%%%%%%
%%%%%%%%%%%%%%%%% AUTHOR INFO. %%%%%%%%%%%%%%%%%
%%%%%%%%%%%%%%%%%%%%%%%%%%%%%%%%%%%%%%%%%%%%%%%%
\noindent\HorRule\color{black}

\medskip \noindent {\bf Acknowledgements.} We thank the anonymous referees for their insightful comments that helped to improve the paper. We thank Thomas M\"{u}ller from the Max Planck Institut f\"{u}r Astronomie for assisting with the data visualisation and production of the Supplementary Videos. We thank Jill Rathborne for making the data on G0.253+0.016 available, and Paola Caselli, Bruce Elmegreen, and Juan Soler for discussions. The `compass' at the centre of Figures 1 \& 2 is adapted from the sketch of the Galaxy produced by Robert Hurt of the Spitzer Science Center (SSC), full credits, NASA/JPL-Caltech/R.~Hurt (SSC/Caltech). J.M.D.K. and M.C. acknowledge funding from the German Research Foundation (DFG) in the form of an Emmy Noether Research Group (grant number KR4801/1-1) and a DFG Sachbeihilfe grant no. KR4801/2-1. J.M.D.K. acknowledges funding from the European Research Council (ERC) under the European Union's Horizon 2020 research and innovation programme via the ERC Starting Grant MUSTANG (grant agreement number 714907). M.R. and J.K. acknowledge funding from the European Union's Horizon 2020 research and innovation program under grant agreement No 639459 (PROMISE). The work of A.K.L. and J.S. is partially supported by the National Science Foundation (NSF) under Grants No.1615105, 1615109, and 1653300, and by NASA under ADAP grants NNX16AF48G and NNX17AF39G. E.R. acknowledges the support of the Natural Sciences and Engineering Research Council of Canada (NSERC), funding reference number RGPIN-2017-03987. C.B. gratefully acknowledges support from the National Science Foundation under Grant No. (1816715). R.S.K. and S.C.O.G. acknowledge funding from the Deutsche Forschungsgemeinschaft (DFG) via the Collaborative Research Center (SFB 881) 'The Milky Way System' (subprojects A1, B1, and B2) and from the Heidelberg Cluster of Excellence {\em STRUCTURES} in the framework of Germany’s Excellence Strategy (grant EXC-2181/1 - 390900948). E.S. acknowledges funding from the European Research Council (ERC) under the European Union’s Horizon 2020 research and innovation programme (grant agreement No. 694343). F.B. acknowledges funding from the European Union's Horizon 2020 research and innovation programme (grant agreement No 726384). J.P. acknowledges funding from the Programme National ``Physique et Chimie du Milieu Interstellaire'' (PCMI) of CNRS/INSU with INC/INP, co-funded by CEA and CNES.  

\medskip \noindent {\bf Author Contributions.} J.D.H., J.M.D.K., and S.N.L. were responsible for the experiment design. J.D.H.\ led the project, carried out the experiment, developed the analysis methods, interpreted the results, and wrote the text, to which J.M.D.K, S.N.L, and M.R. contributed. E.R. created the mock data sets and designed the observational estimator 2-D model test. J.D.H. and M.R. were responsible for data visualisation, assisted by Thomas M\"{u}ller from the Max Planck Institut f\"{u}r Astronomie. Supplementary videos 1-5 were produced by Thomas M\"{u}ller and J.D.H. Supplementary videos 6-10 were produced by M.R. J.D.H. performed the spectral decomposition for all regions except for the GRS data for which M.R. was responsible. All authors contributed to aspects of the data reduction and analysis, the interpretation of the results, and the writing of the manuscript.

\medskip \noindent {\bf Data Availability.} The $^{13}$CO (1-0) data of the Galactic disc are from the Boston University-FCRAO Galactic Ring Survey (GRS). The GRS is a joint project of Boston University and Five College Radio Astronomy Observatory, funded by the National Science Foundation under grants AST-9800334, AST-0098562, AST-0100793, AST-0228993, \& AST-0507657. These data are publicly available at \url{https://www.bu.edu/galacticring/new_data.html}. The N$_{2}$H$^{+}$ ($1-0$) data of the CMZ was obtained using the Mopra radio telescope, a part of the Australia Telescope National Facility which is funded by the Commonwealth of Australia for operation as a National Facility managed by CSIRO. The University of New South Wales (UNSW) digital filter bank (the UNSW-MOPS) used for the observations with Mopra was provided with support from the Australian Research Council (ARC), UNSW, Sydney and Monash Universities, as well as the CSIRO. These data are publicly available at \url{http://newt.phys.unsw.edu.au/mopracmz/data.html}. The ALMA HNCO $4(0,4)-3(0,3)$ data of G0.253+0.016 are from project 2011.0.00217.S (PI J.~Rathborne) and the raw data are publicly available through the ALMA archive (\url{https://almascience.eso.org/alma-data/archive}). All other data that support the findings of this study are available from the corresponding author upon reasonable request.
    
\medskip \noindent {\bf Code Availability.} {\sc ScousePy} and {\sc acorns}, as well as the codes used for our statistical analyses, are are freely available at \url{https://github.com/jdhenshaw}. {\sc GaussPy+} is available at \url{https://github.com/mriener/gausspyplus}. Assistance with this software can be provided by the corresponding author. 

\medskip \noindent {\bf Author Information.} Correspondence and requests for materials should be addressed to J.D.H. (henshaw@mpia.de).

%%%%%%%%%%%%%%%%%%%%%%%%%%%%%%%%%%%%%%%%%%%%%%%%
%%%%%%%%%%%%%%%%% REFERENCES. %%%%%%%%%%%%%%%%%%
%%%%%%%%%%%%%%%%%%%%%%%%%%%%%%%%%%%%%%%%%%%%%%%%

\bibliographystyle{naturemag}
\bibliography{references}

\end{document}